\documentclass[sigconf]{acmart}

\setcopyright{none} 
\acmConference[Anonymous Submission to ACM CCS 2021]{ACM Conference on Computer and Communications Security}{Due 20 January 2021}{London, TBD}
\acmYear{2021}

\usepackage{tikz}
\usepackage{amsmath}
\usepackage{filecontents}
\usepackage{footnote}
\usepackage{diagbox}
\usepackage{listings}
\usepackage{booktabs}
\usepackage{multirow}
\usepackage[ruled,vlined
]{algorithm2e}
\usepackage{subfigure}
\usepackage{tabularx}
\usepackage{url}
\usepackage{amsfonts}
\usepackage{algorithmic}
\usepackage{graphicx}
\usepackage{textcomp}
\usepackage{xcolor}
\usepackage[skins]{tcolorbox}
\usepackage{verbatim}
 
\usepackage{enumitem}
\usepackage{pifont}
\usepackage{listings}
\usepackage[framemethod=tikz]{mdframed}
\usepackage{balance}

\def\BibTeX{{\rm B\kern-.05em{\sc i\kern-.025em b}\kern-.08em
    T\kern-.1667em\lower.7ex\hbox{E}\kern-.125emX}}
\def\eg{\emph{e.g.}\xspace}
\def\etc{\emph{etc.}\xspace}
\def\ie{\emph{i.e.}\xspace}
\def\etal{\emph{et al.}\xspace}

\definecolor{mygreen}{rgb}{0,0.6,0}  
\definecolor{mygray}{rgb}{0.5,0.5,0.5}  
\definecolor{mymauve}{rgb}{0.58,0,0.82}  
\definecolor{lightgrey}{rgb}{0.95,0.95,0.92}
\definecolor{codegreen}{rgb}{0,0.6,0}
\definecolor{codegray}{rgb}{0.5,0.5,0.5}
\definecolor{codepurple}{rgb}{0.58,0,0.82}

\lstdefinestyle{mystyle}{  
    commentstyle=\color{codegreen},
    keywordstyle=\color{magenta},
    numberstyle=\footnotesize\color{codegray},
    stringstyle=\color{codepurple},
    basicstyle=\ttfamily\linespread{1.2},
    breakatwhitespace=false,         
    breaklines=true,                 
    captionpos=b,                    
    keepspaces=false,                 
    numbers=left,                    
    numbersep=5pt,
    xleftmargin=1cm,
    showspaces=false,                
    showstringspaces=false,
    showtabs=false,                  
    tabsize=2,
    frame=tlbr, 
    escapeinside={<@}{@>},
    linewidth=0.9\linewidth
}

\definecolor{mycolor}{rgb}{0.122, 0.435, 0.698}

\begin{document}

 \setcopyright{acmcopyright}
 \copyrightyear{2021}
 \acmYear{2021}
 \acmDOI{10.1145/1122445.1122456}

\acmConference[CCS 2021]{The 28th ACM Conference on Computer and Communications Security}{14 - 21 November, 2021}{Seoul, South Korea}
\acmBooktitle{2021 ACM SIGSAC Conference on Computer and Communications Security (CCS '21), November 14–19, 2021, Virtual Event, South Korea}
\acmPrice{15.00}
\acmISBN{978-1-4503-XXXX-X/18/06}

\settopmatter{printacmref=true, printccs=true, printfolios=true} 

\title{\textsc{Snipuzz}: Black-box Fuzzing of IoT Firmware via\\ Message Snippet Inference} 

\author{Xiaotao Feng$^{\ast}$, Ruoxi Sun$^{\dagger}$, Xiaogang Zhu$^{\ast\ddagger}$, Minhui Xue$^{\dagger}$, \and 
Sheng Wen$^{\ast}$, Dongxi Liu$^{\ddagger}$, Surya Nepal$^{\ddagger}$, Yang Xiang$^{\ast}$}
\affiliation{
\institution{$^{\ast}$Swinburne University of Technology, Australia}
\institution{$^{\dagger}$The University of Adelaide, Australia}
\institution{$^{\ddagger}$CSIRO Data61, Australia}
\country{}
}

\renewcommand{\shortauthors}{X. Feng, R. Sun, X. Zhu, M. Xue, S. Wen, D. Liu, S. Nepal, and Y. Xiang}
 \renewcommand \authors{Xiaotao Feng, Ruoxi Sun, Xiaogang Zhu, Minhui Xue, Sheng Wen, Dongxi Liu, Surya Nepal, and Yang Xiang}

\begin{abstract}

The proliferation of Internet of Things (IoT) devices has made people's lives more convenient, but it has also raised many security concerns. Due to the difficulty of obtaining and emulating IoT firmware, in the absence of internal execution information, black-box fuzzing of IoT devices has become a viable option. However, existing black-box fuzzers cannot form effective mutation optimization mechanisms to guide their testing processes, mainly due to the lack of feedback. In addition, because of the prevalent use of various and non-standard communication message formats in IoT devices, it is difficult or even impossible to apply existing grammar-based fuzzing strategies. Therefore, an efficient fuzzing approach with syntax inference is required in the IoT fuzzing domain.

To address these critical problems, we propose a novel automatic black-box fuzzing for IoT firmware, termed \textsc{Snipuzz}. \textsc{Snipuzz} runs as a client communicating with the devices and infers message snippets for mutation based on the responses. Each snippet refers to a block of consecutive bytes that reflect the approximate code coverage in fuzzing. This mutation strategy based on message snippets considerably narrows down the search space to change the probing messages. We compared \textsc{Snipuzz} with four state-of-the-art IoT fuzzing approaches, \ie, \textsc{IoTFuzzer}, \textsc{BooFuzz}, \textsc{Doona}, and \textsc{Nemesys}. \textsc{Snipuzz} not only inherits the advantages of app-based fuzzing (\eg, \textsc{IoTFuzzer}), but also utilizes communication responses to perform efficient mutation. Furthermore, \textsc{Snipuzz} is lightweight as its execution does not rely on any prerequisite operations, such as reverse engineering of apps. We also evaluated \textsc{Snipuzz} on 20 popular real-world IoT devices. 
Our results show that \textsc{Snipuzz} could identify 5 zero-day vulnerabilities, and 3 of them could be exposed \textit{only} by \textsc{Snipuzz}. All the newly discovered vulnerabilities have been confirmed by their vendors.
\end{abstract}

\maketitle

\section{Introduction}
The Internet of Things (IoT) refers to the billions of physical devices around the world which are now connected to the Internet, all collecting and sharing data. As early as 2017, IoT devices have outnumbered the world's population~\cite{zdnet}, and by 2020, every person on this planet has four IoT devices on average~\cite{lant}. While these devices enrich our lives and industries, unfortunately, they also introduce blind spots and security risks in the form of vulnerabilities. We take Mirai~\cite{trendmicroMirai} as an example. Mirai is one of the most prominent types of IoT botnet malware. In 2016, Mirai took down widely-used websites in a distributed denial of service (DDoS) campaign consisting of thousands of compromised household IoT devices. In the case of Mirai, attackers exploited vulnerabilities to target IoT devices themselves and then weaponized the devices for larger campaigns or spreading malware to the network. In fact, attackers can also use vulnerable devices for lateral movement, allowing them to reach critical targets. For example, in the work-from-home scenarios during COVID-19, Trend Micro has reported that, introducing vulnerable IoT devices to the household will expose employees to malware and attacks that could slip into a company's network~\cite{trendmicro}. Considering the ubiquity of IoT devices, we believe that these known security incidents and risky scenarios are nothing but a tip of the iceberg.

IoT vulnerabilities are normally about the implementation flaws within a device's firmware. To launch new products as soon as possible, developers always tend to use open-source components in firmware development without good update plans~\cite{eclipse}. This sacrifices the security of IoT devices and exposes them to vulnerabilities that security teams cannot remedy quickly. Even if vendors plan to fix the vulnerabilities in their products, the over-the-air patching is usually infeasible because IoT devices do not have reliable network connectivity~\cite{pwnie}. As a result, half of the IoT devices in the market were reported to have vulnerabilities~\cite{lindsey}. 

It is hence crucial to discover such vulnerabilities and fix them before an attacker does. However, most IoT software security tests heavily rely on the assumption of device firmware availability. In many cases, manufacturers tend not to release their product firmware and that makes various dynamic analysis methods based on code analysis~\cite{Costin2014Large,Pewny2015Cross, celik2018soteria,feng2016scalable,Xu2017Neural, Eschweiler2016discovRE} (or emulation~\cite{Chen2016Towards,Clements2020HALucinator,Gustafson2019Analysis,Zheng2019FIRM,Zaddach2014AVATAR}) difficult. Among the existing defense techniques, fuzz testing has shown promises to overcome these issues and has been widely used as an efficient approach in finding vulnerabilities. Moreover, the ability of IoT devices to communicate with the outside world offers us a new option, and that is to test device firmware through exchanging network messages. Therefore, an IoT fuzzer could be designed to send random communication messages to the target device in order to detect if it shows any symptoms of malfunctioning. Potential vulnerabilities could be exposed if crashes are triggered during execution or the device is pushed to send back abnormal messages. 

However, using network communication to fuzz the firmware of IoT devices is very challenging. 
Since obtaining internal execution information from the device is not possible, most existing network IoT fuzzers~\cite{Chen2018IOTFUZZER,pereyda2017boofuzz,doona} work in a black-box manner. This makes optimizing the mutation strategies very difficult. 
Because the selection of mutated seeds is entirely random, existing black-box IoT fuzzing approaches could become very hard to handle, and sometimes, even become more like brute force crack testing. 
In addition, IoT devices have strict grammatical specifications for inputs in communication. Most of the messages that are generated by random mutation will break the syntax rules of the input, and will be quickly rejected during syntax validation in the firmware before being executed. A grammar-based mutation strategy~\cite{wang2017skyfire,aschermann2019nautilus} can effectively generate messages that meet the input requirements though. This can be done by learning the syntax via documented grammatical specifications or from a labeled training set. However, as shown in Table~\ref{tab:devices}, many non-standard IoT device communication formats are being used in practice. 
Therefore, preparing enough learning materials for grammar-based mutation strategies is a huge workload, which makes the deployment of grammar-based IoT fuzzing difficult.

\vspace{1mm}
\noindent\textbf{Challenges.} In this paper, we focus on detecting vulnerabilities in IoT firmware by sending messages to IoT devices. To design an effective and efficient fuzzing method, several  challenges have to be overcome.

\begin{itemize}[leftmargin=*]
    \item \textit{Challenge 1: Lack of a feedback mechanism.} Without access to firmware, it is nearly impossible to obtain the internal execution information from IoT device to guide the fuzzing process (as is done in most typical fuzzers). Therefore, we need a lightweight solution to obtain feedback from device, and optimize the generation process.
    
    \item \textit{Challenge 2: Diverse message formats.} Table 1 shows some message formats that are used in IoT communication, including JSON, SOAP, Key-value pairs, string, or even customized formats. In order to be applied to various devices, a solution should be able to infer the format from a raw message.
    
    \item \textit{Challenge 3: Randomness in responses.} The response messages of an IoT device may contain random elements, such as timestamps or tokens. Such randomness results in different responses for the same message, and diminishes the effectiveness of fuzzing because the input generation of \textsc{Snipuzz} relies on responses.
\end{itemize}

\noindent \textbf{Our approach.} In this paper, we propose a novel and automatic black-box IoT fuzzing, named \textsc{Snipuzz}, to detect vulnerabilities in IoT firmware. 
Different from other existing IoT fuzzing approaches, \textsc{Snipuzz} implements a snippet-based mutation strategy which utilizes feedback from IoT devices to guide the fuzzing.
Specifically, \textsc{Snipuzz} uses a novel heuristic algorithm to detect the role of each byte in the message. It will first mutate bytes in a message one by one to generate probe messages, and categorize the corresponding responses collected from device. Adjacent bytes that have the same role in the message form the initial message snippets, which is the basic unit of mutation. Moreover, \textsc{Snipuzz} utilizes a hierarchical clustering strategy to optimize mutation strategies and reduce the misclassification of categories caused by randomness in the response messages and the firmware's internal mechanism. 
Therefore, \textsc{Snipuzz}, as a black-box fuzzer, can still effectively test the firmware of IoT devices without the support of grammatical rules and internal execution information of the device.

\textsc{Snipuzz} resolves \textbf{Challenge 1} by using responses as the guidance to optimize the fuzzing process. Based on the responses, \textsc{Snipuzz} designs a novel heuristic algorithm to initially infer the role of each byte in the message, which resolves \textbf{Challenge 2}. 
Snipuzz utilizes edit distance~\cite{edit} and agglomerative hierarchical clustering~\cite{cluster} to resolve \textbf{Challenge 3}. We summarize our main contributions as follows:

\begin{table}[t]
    \centering
    \small
    \caption{Format requirements of IoT Devices.}
    \vspace{-3mm}
    \resizebox{\linewidth}{!}{
    \begin{tabular}{rlllll}
    \toprule
    \textbf{\#} & \textbf{Device Type} & \textbf{Vendor} & \textbf{Model} &  \begin{tabular}[c]{@{}l@{}}\textbf{Firmware}\\\textbf{Version}\end{tabular} & \textbf{Format} \\
    \midrule
    1 & Smart Bulb & Yeelight & YLDP05YL & 1.4.2\_0016 & JSON \\
    2 & Smart Bulb & Yeelight & YLDP13YL & 1.4.2\_0016 & JSON \\
    3 & Smart Bulb & Philips & A60 & 1.46.13\_r26312 & JSON \\
    4 & Smart Bulb & LIFX & Mini C & v3.60 & Custom Byte \\ 
    5 & Smart Bulb & FloodLight & BR30 & 35.V7.63.7189-A & Custom Byte \\
    6 & Home Bridge & Philips & Hue & 1935144040 & JSON \\
    7 & Home Bridge & Alro & Base Station & 1.12.2.8\_9\_fc4b603 & JSON \\
    8 & Smart Plug & Tplink & HS100 & 1.5.2 & JSON  \\
    9 & Smart Plug & Tplink & HS110 & 1.5.2 & JSON$^{\ast}$ \\
    10 & Smart Plug & Belkin WeMo & F7C027au & 2.00.1821 & SOAP \\
    11 & Smart Plug & Meross & MSS310 & 2.1.14 & JSON$^{\ast}$ \\ 
    12 & Smart Plug & Orvibo & B25AUS & v3.1.3 & JSON \\ 
    13 & Smart Plug & Konke & Mini US & us1.1.0 & String \\  
    14 & Smart Plug & Broadlink & SP4L-AU & v57209 & Custom Byte \\ 
    15 & Router & Netgear & R6400 & 1.0.1.46 & SOAP$^{\ast}$ \\ 
    16 & TA Assistant & ZKteco & WL10 & ZLM-FX1-3.0.23 & Custom Byte \\
    17 & Camera & Alro & Alro Pro 2 & 1.125.14.0\_34\_1189 & JSON$^{\ast}$ \\
    18 & Camera & Foscam & F19821W & 2.21.1.127 & JSON$^{\ast}$ \\
    19 & NAS & QNAP & T-131P & 4.3.6.0959 & Key-value pairs \\
    20 & Universal Remote & BroadLink & RM mini 3 & v44057 & Custom Byte \\
    \bottomrule
    \end{tabular}
    }
    \begin{flushleft}\footnotesize\quad$^{\ast}$: have randomness in response.\end{flushleft}
    
    \label{tab:devices}
\end{table}

\begin{itemize}[leftmargin=*]
    \item \textbf{Message snippet inference mechanism.} The responses from IoT devices are related to code execution path in firmware. Based on responses, we infer the relationship between message snippets and code execution path in firmware. This novel mutation mechanism enables that \textsc{Snipuzz} does not need any syntax rules to infer the hidden grammatical structure of the input through the device responses. Compared with the actual syntax rules that determine the input string format, the result of snippet determination proposed by \textsc{Snipuzz} has a similarity of 87.1\%.
    
    \item \textbf{More effective IoT fuzzing.} When testing IoT devices, the number of response categories is positively correlated with the number of code execution paths in the firmware. In the experiment, the number of response categories explored by \textsc{Snipuzz} far exceeded other methods on most devices, no matter how long the analysis duration was (in 10 minutes or 24 hours). 
    
    \item \textbf{Implementation and vulnerability findings.}  We implemented the prototype of \textsc{Snipuzz}.\footnote{Publicly available at \url{https://github.com/XtEsco/Snipuzz}.} We used it to test 20 real-world consumer-grade IoT devices while comparing with the state-of-the-art fuzzing tools, \ie, \textsc{IoTFuzzer}, \textsc{Doona}, \textsc{Boofuzz}, and \textsc{Nemesys}. In 5 out of 20 devices, \textsc{Snipuzz} successfully found 5 zero-day vulnerabilities, including null pointer exceptions, denial of service, and unknown crashes, and 3 of them could be exposed \textbf{only} by \textsc{Snipuzz}.
\end{itemize}

\begin{figure*}[t]
\centering

\includegraphics[width=0.85\linewidth]{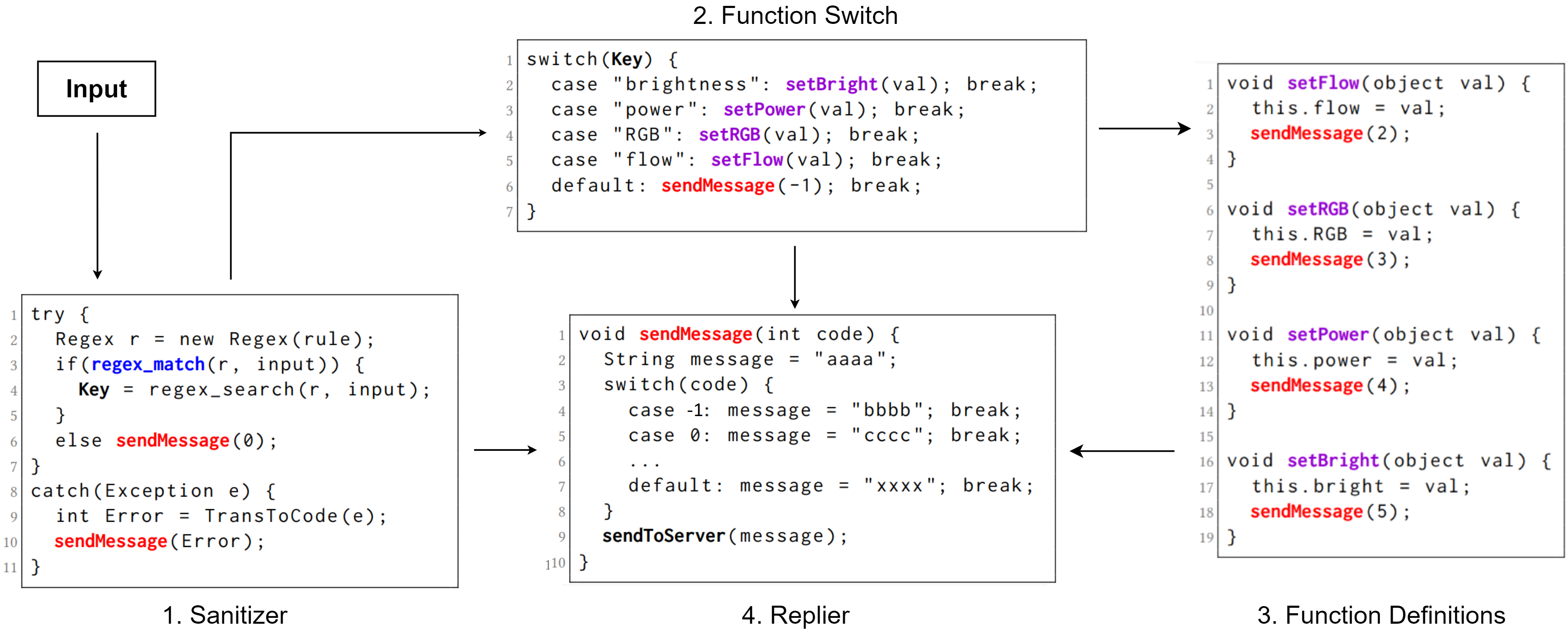}
\caption{\small Interaction with IoT Firmware. Most implementations of IoT devices have a similar communication architecture, including Sanitizer, Function Switch, Function Definitions, and Replier. If the Sanitizer and the Function Switch perform correctly, corresponding functionalities will be executed. Except for crashes, the Replier will always send responses to clients.}\label{fig:structure}
\vspace{-2mm}
\end{figure*}

\section{Background}

\subsection{Fuzz Testing}

Fuzzing is a powerful automatic testing tool to detect software vulnerabilities. After decades of development, fuzzing has been widely used as a base in several security testing domains, such as the OS kernel~\cite{KAFL,corina2017difuze}, servers~\cite{Pham2020AFLNET}, and the blockchain~\cite{gasfuzzer}.

In general, fuzzing feeds the target programs with numerous mutated inputs and monitors exceptions (\eg, crashes). If an execution reveals undesired behavior, a vulnerability could be detected.
To discover vulnerabilities more effectively, fuzzing algorithms optimize the mutation process based on feedback of executions (\eg, coverage knowledge), instead of using a purely random mutation strategy. Moreover, fuzzers can judge from the feedback mechanism whether each test case generated by seed mutation is ``interesting'' (\ie, whether the test case has explored unseen execution states). If a test case is interesting, it will be reserved as a new seed to participate in future mutation. 
With the feedback, many fuzzers~\cite{Boehme2017Directed,Boehme2016Coverage,Wang2020Not,Oesterlund2020ParmeSan,Yue2020EcoFuzz} steer the computing resources towards the interesting test cases and achieve higher possibility to discover vulnerabilities.

\subsection{Generic Communication Architecture of IoT Devices}

To react with external inputs, most IoT devices implement a similar high-level communication architecture. As per the pseudo code example presented in Figure~\ref{fig:structure}, a typical implementation of the communication architecture may consist of four parts: 1) Sanitizer, 2) Function Switch, 3) Function Definitions, and 4) Replier.

When an IoT device receives an external input, Sanitizer starts parsing the input and performs regular matching. If the input format breaches the syntactic requirements, or an exception occurs during the parsing process, Sanitizer will directly notify Replier by sending a response message describing the input error and terminate the processing of input.
If the input is syntactically correct, Function Switch transfers control to the corresponding Functions according to the attribute, \texttt{Key}, and corresponding value, \texttt{val}, extracted from the input. If \texttt{Key} cannot be matched, the processing of this input will be terminated, similarly as done by Replier. 
When Functions completes the processing, such as \texttt{setFlow()}, with the parameter \texttt{val}, it notifies Replier to generate the response message. Note that, the implementation of Functions is specific to IoT devices. 
As described above, Replier is responsible for sending responses to the client (such as the user's APP). 
Based on the calling situation (indicated by the parameter \texttt{code} in the example), Replier determines the content of response message to be sent. 

\section{Motivation}

\subsection{Response-Based Feedback Mechanism}

The interactive capabilities of IoT devices make it possible to test security of device firmware through the network. However, there are also some challenges when testing IoT devices using network-based fuzzers. Since most network fuzzing methods cannot directly obtain execution status of the device, it is hard to establish an effective feedback mechanism to guide the fuzzing process. Without feedback mechanism, the fuzzing tests could be blind in the selection of mutation targets, and may lean to a brute force random test.

As discussed previously, due to the lack of open-sourced firmware, it is difficult or even impossible to instrument the IoT devices. Therefore, the response messages returned by the firmware can be regarded as a valuable source of device status information at run-time.
The Replier in Figure~\ref{fig:structure} will use the value of the variable \texttt{code} to determine the content of the response messages. The value of \texttt{code} comes from many different function blocks in the firmware. Parameters are passed when Sanitizer fails to parse the input or some exceptions are triggered; or when the Function Switch cannot match the key command characters in the input; or after each input is executed in the Functions. Therefore, through the content of the response message, the code block that has been executed in the firmware can be inferred. 
When the firmware source code is not available, the correspondence between the firmware execution and the response messages cannot be directly extracted. Moreover, the firmware may return the same response messages even executing different functions.

Although the response message cannot be equated to the execution path of the device, it can still play an important role in the black-box fuzz testing for IoT devices. Although it is hard to link the code execution path corresponding to each response message, if the two inputs get different response messages, we can deduce that the two inputs go to different firmware code execution paths.

\noindent\textbf{Our approach.} \textsc{Snipuzz} uses the response message to establish a new feedback mechanism. \textsc{Snipuzz} will collect every response, and when a new response is found, the input corresponding to the response will be queued as a seed for subsequent mutation testing.

\subsection{Message Snippet Inference} \label{movti_infer}

The firmware of the IoT device can be regarded as a software program with strict syntax requirements for input. If the byte-based mutation strategies (such as mutating each byte in the input one by one or randomly selecting bytes for mutation testing) are used in the fuzz testing, the generated test cases could be rare to meet the input syntax requirements. The grammar-based fuzzers utilize detailed documents or a large training data set to learn the grammatical rules and use it to guide the generation of mutation~\cite{wang2017skyfire, pham2019smart}. 
In many cases, the input syntax in IoT devices is diverse or non-standard. Table~\ref{tab:devices} shows the communication format requirements used in 20 IoT devices from different vendors. Some of them are using well-known formats such as JSON and SOAP, but some use Key-value pairs or even custom strings as communication format. Therefore, it is difficult to provide grammar specifications or establish training data sets that cover communication formats on a large scale for the grammar-based mutation strategy.

\textit{The best grammar guidance originates from the firmware itself.}
Responses from IoT devices suggest the execution results of messages. 
If we mutate a valid message byte by byte (\ie, breaching the format), we will get many different responses. If mutation of two different positions in the valid message receives the same response, these two positions have a high possibility that they are related to the same functionality in firmware.
Therefore, those consecutive bytes with the same response can be merged into one snippet. This method of inferring message snippets can clearly reflect the utility of each byte after entering the firmware.
In addition, mutation based on message snippets can largely reduce the search space and improve the efficiency of fuzzing.

\noindent\textbf{Our approach.} \textsc{Snipuzz} merges consecutive bytes with the same response into one snippet. We also propose different mutation operators performing on snippets.

\section{Methodology}

\begin{figure*}[t]
    \centering
    \includegraphics[width=0.95\textwidth]{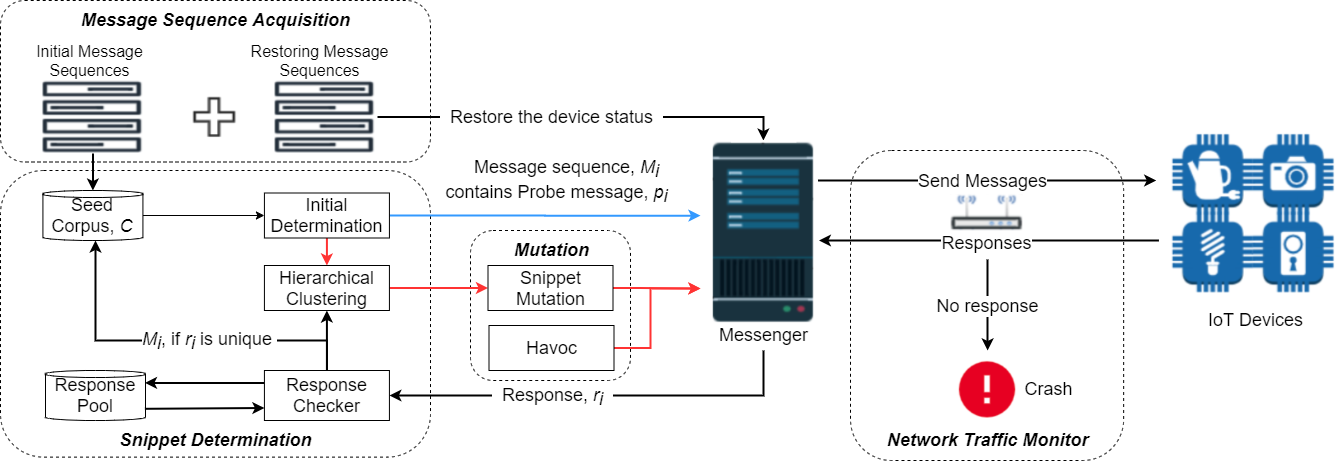}
    \caption{\small Workflow of \textsc{Snipuzz}. With the valid message sequences (seeds), \textsc{Snipuzz} performs snippet determination on each individual message. Then, \textsc{Snipuzz} mutates snippet(s) to generate new message sequences. By monitoring the network traffic, \textsc{Snipuzz} determines a crash when no responses are received.}
    \label{fig:workflow}
    \vspace{-2mm}
\end{figure*}

In order to clearly present our approach, we first introduce some notations while explaining the fuzzing process of \textsc{Snipuzz}. At a high level, \textsc{Snipuzz} performs as a client which sends a  \textit{\textbf{message sequence}} $M$ to request certain actions from IoT devices. Any message $m \in M$ requests the IoT device to perform a certain functionality, and all the messages $\bigcup\limits_k m_k = M$ work together to request an action (or actions). 
Similarly to the typical fuzzers, we initialize a \textit{\textbf{seed}} $S$ with an \textit{\textbf{initial message sequence}}, and a seed corpus $C$ with all the seeds (Section~\ref{sec:initial_seed_acquisition}). Meanwhile, \textit{\textbf{restoring message sequences}} are collected for resetting the IoT device to a predefined status.

To establish an effective fuzzing, as depicted in Figure~\ref{fig:workflow}, \textsc{Snipuzz} first conducts a snippet determination process. Concretely, \textsc{Snipuzz} selects a message $m$ in a seed $S \subset C$, from which a \textit{\textbf{probe message}} $p_{i}$ and a corresponding sequence $M_{i}$ will be generated. 
Each message in $M_{i}$ will trigger a \textit{\textbf{response message}} $r_{i}$ (response for short) containing the information about the execution output. \textsc{Snipuzz} assigns each message $m$ a \textit{\textbf{response pool}} $R$, which is utilized to determine if a new response $r_{i}$ is unique. The uniqueness of a response indicates that it does not belong to any category of responses existed in the response pool. 
If $r_{i}$ is unique, \textsc{Snipuzz} will add $r_{i}$ into the pool $R$, and reserve the corresponding message sequence $M_{i}$ as a new seed.
\textsc{Snipuzz} then divides the message $m$ into different snippets based on the responses (Section~\ref{sec:snippet_determination}).
Upon the snippets are obtained, \textsc{Snipuzz} performs mutation according to various strategies, \eg, empty, bytes flip, data boundary, or havoc (detailed in Section~\ref{mutations}). Throughout the fuzzing process, \textsc{Snipuzz} sets up a network monitor to detect crashes which may indicate vulnerabilities (Section~\ref{net_monitor}).

\subsection{Message Sequence Acquisition} \label{sec:initial_seed_acquisition}

The quality of initial seeds could influence the fuzzing campaigns significantly. 
Therefore, we consider to obtain high-quality initial seeds conforming to highly-structured formats required by IoT devices, as such inputs may exercise complex execution paths and enlarge the opportunity of exposing vulnerabilities at deep code.
Generating seeds based on companion app reverse-engineering~\cite{Chen2018IOTFUZZER} or accessible specifications (as mentioned in Section~\ref{movti_infer}) could be intuitive solutions. However, they either require heavy engineering efforts or could be error-prone (\eg, seeds may violate the required formats or have the wrong order of messages). 

\noindent\textbf{Initial seed acquisition}. \textsc{Snipuzz} proposes a lightweight solution to obtain initial valid seeds. Considering that many IoT devices have first- or third-party API documents as well as the test suites, the testing programs provided by both parties can effectively act as a client, sending control commands to IoT devices or remote servers. Most structural information (\eg, header, message content) and protocols (\eg, HTTP, HNAP, MQTT) of communication packets are defined in the API programs as message payloads. 
Therefore, \textsc{Snipuzz} leverages these test suites to communicate with the target devices, while at the same time, extracting the message sequences as initial seeds.
For example, when using an API program to turn on a light bulb, the program first sends login information to the server or to the IoT device, then sends a message to locate a specific light bulb device, and finally sends a message to control the device to turn on the light. \textsc{Snipuzz} captures such a message sequence that triggers a functionality of IoT device as an initial seed.

\noindent\textbf{Restoring message sequence acquisition}. In order to replay a test case for the crash triage, \textsc{Snipuzz} ensures that the device under test has the same initial state in each round of testing. After sending any message sequence to the device, \textsc{Snipuzz} will send a restoring message sequence to reset the device to a predefined status. 

\noindent \textbf{Manual efforts}. Although we try our best efforts to provide a lightweight fuzzer, \textsc{Snipuzz} still requires some manual efforts to obtain valid and usable initial seeds. First, we manually configure the programs from the test suites, such as setting up the IP address and the login information. Note that, we only need to configure these programs once per device.
Second, to capture the message sequences dynamically, we need to manually define the specific format and protocol in the network traffic monitor. 
Finally, we filter out some message sequences that will mislead the fuzzing process. 
For instance, some API programs provide operations that can automatically update or restart the device. These operations will halt the device and thus no response will be sent back. This leads to false-positive crashes because we consider a no-response execution as a crash. 
The manual work costs roughly 5 man-hours per device and is only required during the message sequence acquisition phase of \textsc{Snipuzz}.

\subsection{Snippet Determination}\label{sec:snippet_determination}

The key idea of \textsc{Snipuzz} is to optimize fuzzing process based on snippets determined by responses. Put differently, \textsc{Snipuzz} leverages snippet mutation to reduce the search space of inputs, while the snippets are automatically clustered via categorizing responses from IoT devices. The major challenge is to correctly understand the semantics of responses. For instance, due to the presence of timestamp, two semantically identical responses will be classified into different categories if utilizing a simple string comparison. 
Therefore, \textsc{Snipuzz} utilizes a heuristic algorithm and a hierarchical clustering approach to determine the snippets in each message.

\subsubsection{Initial Determination}\label{sec:SD}

The essence of a message snippet is the consecutive bytes in a message that enables the firmware to execute a specific code segment.
For experienced experts, it is not difficult to segment message snippets according to the semantic definition in official documents.
However, for algorithms that lack such knowledge, it is essential to apply some automatic approaches to identify the meaning of each byte in the message.

\textsc{Snipuzz} first uses a heuristic algorithm to roughly divide each message into initial snippets. The core idea of the heuristic algorithm is to generate \textit{probe messages} $p_{i}$ by deleting a certain byte in the message $m$ ($m \in seed~S$). 
By categorizing the responses $r_{i}$ of each probe message, \textsc{Snipuzz} preliminarily determines the snippets in the message $m$.

\begin{table*}[t]
\centering
\caption{Examples of probe messages and corresponding response messages.}
\label{tab:message_exmple1}
\vspace{-3mm}
\small
\begin{tabular}{@{}ll|ll|l@{}}
\toprule
\textbf{Messages} & \textbf{Content} & \textbf{Responses} & \textbf{Content} 
& \textbf{Category}
\\
\midrule
Message $m$ & \{"\textcolor{red}{on}":true\} & Response $r_{0}$ & \{"success":{"/lights/1/state/on":true\}} 
& 0
\\ 
Probe message $p_{1}$ &   "on":true\}  & Response $r_{1}$  
& \{"error":\{"type":2,"address":"/lights/1/state","description":"body contains invalid json"\}\} 
& 1
\\ 
Probe message $p_{2}$ &   \{on":true\}  & Response $r_{2}$  
& \{"error":\{"type":2,"address":"/lights/1/state","description":"body contains invalid json"\}\} 
& 1
\\ 
Probe message $p_{3}$ &   \{"\textcolor{red}{n}":true\}  & Response $r_{3}$  
& \{"error":\{"type":6,"address":"/lights/1/state/\textcolor{red}{n}","description":"parameter, \textcolor{red}{n}, not available"\}\} 
& 2
\\ 
Probe message $p_{4}$ &   \{"\textcolor{red}{o}":true\}  & Response $r_{4}$  
& \{"error":\{"type":6,"address":"/lights/1/state/\textcolor{red}{o}","description":"parameter, \textcolor{red}{o}, not available"\}\} 
& 3
\\ 
Probe message $p_{5}$ &   \{"on:true\}  & Response $r_{5}$  
& \{"error":\{"type":2,"address":"/lights/1/state","description":"body contains invalid json"\}\} 
& 1
\\ 
Probe message $p_{11}$ &   \{"on":true  & Response $r_{11}$  
& \{"error":\{"type":2,"address":"/lights/1/state","description":"body contains invalid json"\}\} 
& 1 \\
\bottomrule

\end{tabular}
\end{table*}

For example, as shown in Table~\ref{tab:message_exmple1}, to determine snippets in the message $m = \underline{\textit{\{"on":true\}}}$, \textsc{Snipuzz} generates \textit{probe messages} by removing the bytes in $m$ one by one. When the first byte ‘\{’ in $m$ is deleted, the corresponding probe message $p_{1}$ is \underline{\textit{"on":true\}}}. 
Similarly, when the second byte is deleted, the corresponding probe message $p_{2}$ is \underline{\textit{\{on":true\}}}. Therefore, the message $m$ with 11 bytes can generate 11 different probe messages ($p_{1}$ to $p_{11}$). 
\textsc{Snipuzz} will send the 11 corresponding message sequences ($M_{1}$ to $M_{11}$) containing the probe messages to the device and collect responses.

\textsc{Snipuzz} then distinguishes the snippets in the message $m$ by categorizing the responses. 
Specifically, the consecutive bytes with the same corresponding response type are merged into the same snippet.
According to the examples illustrated in Table~\ref{tab:message_exmple1}, the Response $r_{1}$, $r_{2}$, and $r_{5}$ are merged into one category that indicates an error in JSON syntax, while Response $r_{3}$ and $r_{4}$ are merged into another category which indicates an error of an invalid input parameter.
Therefore, the consecutive bytes whose corresponding responses belong to the same category can form a message snippet. 
Through this heuristic approach, \textsc{Snipuzz} can determine all initial snippets in the message $m$.

A naive method to categorize responses is to utilize a string comparison, \ie, comparing the content of responses byte by byte. However, due to the existence of randomness in responses (\eg, timestamp and token), a simple string comparison may incorrectly distinguish the responses with same semantic meaning into different categories. Therefore, a more advanced solution, Edit Distance~\cite{edit}, is introduced to determine the category of responses. As shown in Equation~(\ref{equ:similarity}), a similarity score, $s_{kt}$, between two responses $r_k$ and $r_t$ is calculated.

\begin{equation}
\begin{split}
    s_{kt} = 1 - \frac{edit\_distance(r_k, r_t)}{max\_len(r_k, r_t)},
\end{split}
\label{equ:similarity}
\end{equation}

\noindent where the $max\_len()$ in the equation selects the longer string between the two responses and the $edit\_distance()$ counts the minimum number of operations, including insertion, deletion, and substitution, required to transform one string into the other.
Therefore, the more similar two responses are, the larger the value of $s_{kt}$ is.

\textsc{Snipuzz} first calculates a self-similarity score $s_{ii}$ for each probe message $p_{i}$. 
Note that $p_{i}$ is generated by mutating the $i$-th byte in the message $m$.
Concretely, \textsc{Snipuzz} sends the same probe message $p_{i}$ twice within an interval of one second. 
Two responses $r_{i}, r'_{i}$ will be collected from the IoT device, correspondingly.
The self-similarity score $s_{ii}$ is then calculated based on the two responses $r_{i}, r'_{i}$ according to Equation (\ref{equ:similarity}). Note that, due to the randomness in the responses, there could be differences between the two responses $r_{i}, r'_{i}$, even though they are from the same probe message. Therefore, the self-similarity score could be smaller than 1.

To determine whether two responses belong to the same category, \textsc{Snipuzz} computes the similarity score of two responses and compares it with the self-similarity score. 
For example, for two responses $r_{i}$ and $r_{j}$, \textsc{Snipuzz} uses the Equation~(\ref{equ:similarity}) to compute the similarity score $s_{ij}$. After that, $s_{ij}$ will be compared with the self-similarity. If $s_{ij} >= s_{ii}$ or $s_{ij} >= s_{jj}$ satisfies, responses $r_{i}$ and $r_{j}$ will be considered belonging to the same category; otherwise, responses $r_{i}$ and $r_{j}$ are then assigned to the different categories. 

For a newly received response $r_{i}$, \textsc{Snipuzz} will compare it with all the responses in the corresponding response pool $R$ based on the similarity score. 
If the new response $r_{i}$ does not belong to any existing category, the response $r_{i}$ as well as the corresponding probe message $p_{i}$ will be added into the \textit{Response Pool}.

\begin{figure}[t]
\centering
\includegraphics[width=0.6\linewidth]{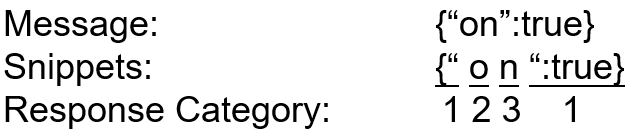}
\caption{\small An example of snippet determination.}\label{fig:snippet_example}
\end{figure}

With the response pool $R$, \textsc{Snipuzz} categories each byte in the message $m$. Specifically, the category of the $i$-th byte in message $m$ is assigned according to the category of response $r_{i}$. Then the consecutive bytes with the same category will be merged into one snippet.
Figure~\ref{fig:snippet_example} shows an example of the initial snippet determination on the message $m = \underline{\textit{\{"on":true\}}}$ according to the response categories in Table~\ref{tab:message_exmple1}.

\subsubsection{Hierarchical Clustering.}\label{sec:clustering}

Although \textsc{Snipuzz} utilizes similarity comparison to mitigate the mis-categorization caused by randomness in responses, two semantically identical responses may still be mis-categorized into different categories. This could occur when the responses contain contents extracted or copied from probe messages. 
For example, due to the quotation of specific error contents from probe messages, the heuristic algorithm will not assign them to one category. Specifically, the similarity score $s_{34}$ of $m = \underline{\textit{\{"on":true\}}}$ in Table~\ref{tab:message_exmple1} is 0.979, which is smaller than the self-similarity scores $s_{33} = 1.000$ and $s_{44} = 1.000$ (as there is no randomness in the responses). However, these two responses are semantically identical and should be identified into one category, \ie, they are both error messages, indicating parameter syntax errors are located in the probe messages and the device is executing the same code block. 

In order to solve the aforementioned problem, \textsc{Snipuzz} uses agglomerative hierarchical clusters to refine message snippets. The core idea of hierarchical clustering is to continuously merge the two most similar clusters until only one cluster remains. 

As shown in Algorithm~\ref{alg:cluster}, \textsc{Snipuzz} will initialize the snippets according to Initial Snippets determined in Section~\ref{sec:SD} (line 1). After that, each response category in the response pool will be initialized as a cluster (line 2). \textsc{Snipuzz} will convert the responses into feature vectors (line~3, detailed in the later paragraph) which will be used to compute the distance between each pair of clusters (lines 5-7). Then the two closest clusters will be merged and the cluster center will be updated accordingly (lines 8-10). After performing the cluster process, \textsc{Snipuzz} will generate new snippets according to the current cluster result and add the new snippets into the snippet segmentation result (line 11), which will be further used for mutation.

\begin{algorithm}[t]
\SetAlgoLined
\small
\KwIn{Initial Snippets $F_0$, Response Pool $R$}
\KwResult{Snippets $F$}
\nl $F \gets F_0$\;
\nl $C \gets categorize(F_0)$\;
\nl $V \gets vectorize(R)$\;
\nl \While{size($C$) > 1}{
    \nl \For{$i \gets size(C)$ \KwTo $2$}{
        \nl \For{$j \gets size(C)$-1 \KwTo $1$}{
            \nl $D \gets distance_{ij} = \left\lVert v_i - v_j \right\rVert$\;
        }
    }
    \nl $i,j = {argmin}(D)$\;
    \nl $C \gets merge\_cluster(C,i,j)$\;
    \nl $V \gets update\_cluster\_center(V,i,j)$\;
    \nl $F \gets F + generate\_snippets(C)$\;
}
\caption{\small Hierarchical Clustering for Snippets}\label{alg:cluster}
\end{algorithm}

Concretely, \textsc{Snipuzz} first extracts features from responses, which vectorize responses into tuples of the self-similarity score, the length of the response, the number of alphabetic segments, the number of numeric segments, and the number of symbol segments. 
Each segment consists of consecutive bytes that have the same type. For instance, $``123"$ is 1 numeric segment, and there are 2 alphabetic segments and 1 numeric segment in $``a1b"$. 
More specifically, the $r_{1}$ in Table~\ref{tab:message_exmple1} will be vectorized to $v_1 = (1, 91, 10, 2, 10)$. Similarly, responses $r_{3}$ and $r_{4}$ will be converted to $v_2 = (1, 94, 11, 2, 13)$ and $v_3 = (1, 94, 11, 2, 13)$.

\begin{figure}[t]
\centering
\includegraphics[width=0.75\linewidth]{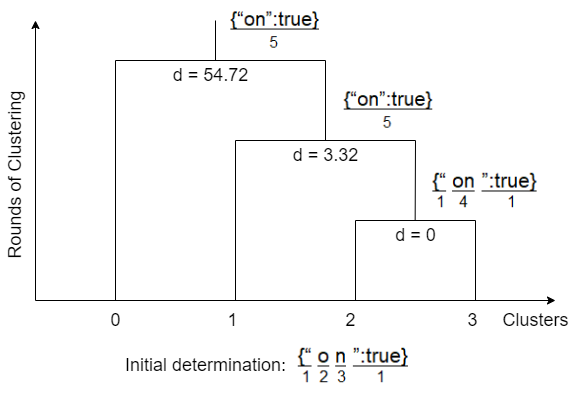}
\vspace{-5mm}
\caption{\small An example of hierarchical clustering. }\label{fig:clustering}
\end{figure}

Figure~\ref{fig:clustering} shows an example of clustering according to the message $m = \underline{\textit{\{"on":true\}}}$ in Table~\ref{tab:message_exmple1}. 
According to the Algorithm~\ref{alg:cluster}, in the preparation round (0th round) of clustering, each category in the response pool will be initialized a single cluster.
In the 1st round, as clusters $2$ and $3$ are the two clusters with minimum distance ($\left\lVert v_2 - v_3 \right\rVert=0$), the two clusters are merged into a new cluster. 
Correspondingly, the message snippets `o' and `n' are merged into a new snippet, marked with index \#4. 
Similarly, in the next round, the two closest clusters, the cluster 1 and the new cluster, are merged, and a new snippet will also be generated. Finally, all snippets in the message are merged into one new snippet, \ie, the message itself. 
All the new generated snippets together with the initial snippets will be used in message mutation in the next stage.

\subsection{Mutation Schemes}\label{mutations}

\noindent \textbf{Snippet Mutation.}
In order to conduct an efficient fuzzing, \textsc{Snipuzz} mutates the snippets obtained in the stage of Snippet Determination. 
Note that the mutation schemes are performed on the entire snippet instead of a single byte in a message. 

\begin{itemize}[leftmargin=*]
\item \textbf{Empty.}
The empty of a data domain may crash the firmware if the data domain is not properly checked. Therefore, \textsc{Snipuzz} deletes an entire snippet to empty the data domain.

    \item \textbf{Byte Flip.}
    To detect bugs in both the syntax parsers and the functional code, \textsc{Snipuzz} flips all bytes in a snippet. This changes the syntactic meaning of strings and will discover bugs when the parser does not properly check syntax. On the other hand, Byte Flip changes the values of data domains to examine firmware.

    \item \textbf{Data Boundary.}
To detect the out-of-bound bugs that occur during assignment, \textsc{Snipuzz} modifies the values of numeric data to some boundary values (\eg, 65536).

    \item \textbf{Dictionary.}
    For the scheme of Dictionary, \textsc{Snipuzz} replaces a snippet with a pre-defined string such as ``true'' and ``false'', which may directly explore more code coverage.

    \item \textbf{Repeat.}
    In order to detect bugs in syntax parsers, \textsc{Snipuzz} repeats a snippet for multiple times. Meanwhile, the repetition of data domain can detect defects caused by out-of-boundary problems.
    
\end{itemize}

\noindent\textbf{Havoc.}
The conditions for triggering bugs may be complicated. For example, it may require modifying different data domains in the same message to trigger a bug. The aforementioned snippet mutation schemes only mutate one snippet at a time. However, the havoc mutation randomly selects some random snippets in a message, and performs the aforementioned mutation schemes on each of the selected snippets. Havoc mutation will not stop until finding a new response category or the target IoT device crashes. 

\subsection{Network Traffic Monitor}\label{net_monitor}

The network communication of the device is monitored and a timeout is set to determine whether the device has been crashed. 
In fact, the monitoring of device network communication is not a single step, and it occurs during the entire fuzzing process. 
In case of timeout, \textsc{Snipuzz} will continue to send the same message sequence for three times, as the cause of timeout could be network fluctuations instead of device crashes. 
If the timeout occurs for three times, \textsc{Snipuzz} will use the control command to physically restart the device and send the same sequence of messages to the device again. If the device still does not return the message on time, \textsc{Snipuzz} will record the crash and the corresponding message sequence.

\subsection{Implementation}

The design of \textsc{Snipuzz} consists of four steps: \textbf{Message Sequence Acquisition}, \textbf{Snippet Determination}, \textbf{Mutation}, and \textbf{Network Communication Monitoring}. In the \textbf{Message Sequence Acquisition} step, we use WireShark~\cite{wireshark} in the program to detect and record the communication packets between the API and the IoT device, and manually cleaned these message sequences. The remaining core functional steps are packaged in a prototype implemented with 4,000 lines of C\# code. The network monitor will record every message sent to the device, and send the information to the device again when the device does not reply. A smart plug was used to implement the physical restart function of the target device. When \textsc{Snipuzz} needs to physically restart the device under test, it will send control messages to the smart plug, and the plug will be closed and then opened. In this way, the device under test will be powered off briefly and restarted.

\section{Experimental Evaluation}\label{sec:experimental_evaluation}
\subsection{Experiment Setup}\label{sec:experiment_setup}

\noindent \textbf{Environment setup.}
To initialize IoT devices, we use the applications provided by the manufacturers to complete the pairing. In order to better monitor the network communication, all devices under test are connected to a local router. Our  automatic packet extractor and \textsc{Snipuzz} run on a Windows 10 desktop PC with Intel Core i7 six-core x 3.70 GHz CPU and 16 GB RAM. The PC is also connected to the router. 

\vspace{1mm}\noindent \textbf{IoT Devices under test.}
We have selected 20 popular consumer IoT devices from both online and offline markets worldwide, covering various well-known brands, such as Philips, Xiaomi, TP-Link, Netgear. The types of selected IoT devices include smart plugs, smart bulbs, routers, home bridge, IP camera, fingerprint terminal, \etc These devices are either recommended items in Amazon or the best-selling products that can be bought in supermarkets. Table~\ref{tab:devices} details the information of the IoT devices under test.

\vspace{1mm}
\noindent \textbf{Benchmark tools.}
In order to verify \textsc{Snipuzz}'s performance in finding crashes and message segmentation, we used seven different fuzzing schemes as benchmarks. 

\begin{itemize}[leftmargin=*]
    \item \textbf{\textsc{IoTFuzzer}~\cite{Chen2018IOTFUZZER}.}
    The core idea of \textsc{IotFuzzer} is to find the functions that send control commands to the IoT device by static analysis of companion apps, and to mutate the value of specific variables to perform fuzzing test without breaking the message format. 
    Note that our implementation of \textsc{IoTFuzzer} is the best effort to replicate since their code is not publicly available, and we acknowledge that this could provide slightly different results with respect to the original version. 
    
    We implement the \textsc{IoTFuzzer} by replacing the mutation algorithm in \textsc{Snipuzz} framework with the mutation strategies in \textsc{IoTFuzzer}. 
    Considering that the purpose of companion apps analysis in \textsc{IoTFuzzer} is to ensure that only the data domain in the communication message is mutated, to make the benchmark as fair as possible, we use seeds same as the ones used in \textsc{Snipuzz} and manually segment the data domain of each seed message before feeding it to \textsc{IoTFuzzer}. We believe that such manual segmentation is sufficient to provide an upper bound performance of \textsc{IoTFuzzer}.
    Note that we remove the methods that are related to the feedback mechanism and snippet segmentation because these methods are not used in \textsc{IoTFuzzer}.  
    
    \item \textbf{\textsc{Nemesys}~\cite{kleber2018nemesys}.}
    \textsc{Nemesys} is a protocol reverse engineering tool for network message analysis. It utilizes the distribution of value changes in a single message to infer the boundaries of each data domain.
    Considering that \textsc{Nemesys} is a protocol inference method instead of an off-the-shelf fuzzing tool, we implement the method of \textsc{Nemesys} based on the \textsc{Snipuzz} framework to infer the snippet boundary, replacing corresponding snippet determination method (Section~\ref{sec:snippet_determination}). 
    \item \textbf{\textsc{BooFuzz}~\cite{pereyda2017boofuzz}.}
    As a successor of \textsc{Sulley}~\cite{sulley}, \textsc{BooFuzz} is an excellent network protocol fuzzer that has been involved in several recent fuzzing research~\cite{Chen2018IOTFUZZER,song2019spfuzz,SGPFuzzer}. 
    Different from other automatic fuzzers, \textsc{BooFuzz} requires human-guided message segmentation strategies as inputs. In our research, we leverage this property and manually define more fuzzing strategies to enrich the benchmark evaluation.
    \begin{itemize}
        \item \textbf{BooFuzz-Default}. In this strategy, we set each message in the input as a complete string, that is, \textsc{BooFuzz} will use the message as a string for mutation testing.
        \item \textbf{BooFuzz-Byte}. 
        Each byte of the message in the input will be used for a mutation test individually.
        \item \textbf{BooFuzz-Reversal}. Contrary to the idea of \textsc{IoTFuzzer}, in this strategy, we focus on the mutation of non-data domain in the message, while keeping data domain unchanged.
    \end{itemize}
    \item \textbf{\textsc{Doona}~\cite{doona}}.  
    \textsc{Doona} is a fork of the Bruterforce Exploit Detector (BED)~\cite{bed}, which is designed to detect potential vulnerabilities related to buffer and formats in network protocol. 
    Different from other tools, \textsc{Doona} does not take network communication packets as seeds. The test cases of \textsc{Doona} are required to be pre-defined for each device or protocol under test.
    \item \textbf{\textsc{Snipuzz-NoSnippet}}.
    \textsc{Snipuzz} uses the segmentation of message snippets to enhance the efficiency of fuzzing and the ability to find crashes. In order to verify whether the snippet determination indeed benefits fuzzing, we implement \textsc{Snipuzz-NoSnippet} based on \textsc{Snipuzz}. \textsc{Snipuzz-NoSnippet} does not have the component of snippet determination, and blindly mutates bytes in messages without the knowledge of responses.

\end{itemize}

{Except for \textsc{Doona}, whose test cases are preset, all benchmark tools and \textsc{Snipuzz} are tested on same input sets. These input sets may be in different formats (\eg, \textsc{BooFuzz} requires to manually set the input, and \textsc{Numesys} requires the input to be the pcap file format), but the content is the same.}

There are many other popular fuzzing tools which are able to test IoT devices via network communication, such as \textsc{Peach}~\cite{Peachtech} and \textsc{AFLNET}~\cite{Pham2020AFLNET}. However, since they are grey-box fuzzing that requires to instrument firmware, it is infeasible and unfair to regard those tools as baselines for black-box schemes.

\subsection{Vulnerability Identification}\label{sec:vuln_identification}

\subsubsection{\textsc{Snipuzz}}

After performing fuzz testing using \textsc{Snipuzz} on each of the 20 IoT devices for 24 hours, we detected 13 crashes in 5 devices. As shown in Table~\ref{tab:result}, the detected crashes include 7 null pointer dereferences, 1 denial of service, and 5 unknown crashes that we further manually verified.
The 13 crashes found by \textsc{Snipuzz} are triggered by providing malformed inputs. These malformed inputs break the message format in different ways. For example, deleting placeholders, emptying the data domain or fortunately changing the type of data value.

Note that all the crashes identified by \textsc{Snipuzz} are in JSON-based devices, although we successfully conducted experiments on the 20 IoT devices with various communication formats, such as JSON, SOAP, and K-V pair. The experiments also show that \textsc{Snipuzz} observes a higher number of response categories compared to the other fuzzers (as detailed in Section~\ref{sec:runtime_perf}).

\begin{table*}[t]
\centering
\footnotesize
\caption{Experiment Results. \textsc{Snipuzz} discovers the most number of categories and exposes the most number of bugs.}
\vspace{-3mm}
\label{tab:result}
\resizebox{\linewidth}{!}{
\begin{tabular}{r|c|ccc|ccc|cc|cc|cc|cc|cc|cc}
\toprule
\multirow{2}{*}{\textbf{\#}} 
& \multirow{2}{*}{\textbf{Devices}} 
& \multicolumn{3}{c|}{\textbf{\textsc{Snipuzz}}} 
& \multicolumn{3}{c|}{\textbf{\textsc{IoTFuzzer}}} 
& \multicolumn{2}{c|}{\textbf{\textsc{Doona}}} 
& \multicolumn{2}{c|}{\textbf{\textsc{BooFuzz-Default}}} 
& \multicolumn{2}{c|}{\textbf{\textsc{BooFuzz-Byte}}} 
& \multicolumn{2}{c|}{\textbf{\textsc{BooFuzz-Reversal}}} 
& \multicolumn{2}{c|}{\textbf{\textsc{Nemesys}}} 
& \multicolumn{2}{c}{\textbf{\textsc{Snipuzz-NoSnippet}}} \\ \cline{3-20}

 & & \textbf{T} & \textbf{C} & \textbf{10/24}
 & \textbf{T} & \textbf{C} & \textbf{10/24}
 & \textbf{C} & \textbf{10/24}
 & \textbf{C} & \textbf{10/24}
 & \textbf{C} & \textbf{10/24}
 & \textbf{C} & \textbf{10/24}
 & \textbf{C} & \textbf{10/24}
 & \textbf{C} & \textbf{10/24}\\
\midrule
1 & YLDP05YL & UC & 3$^{\ast}$ & \textbf{46}/\textbf{71} & UC & 1$^{\ast}$ & 31/33 & NA & NA/NA & 0 & 11/17 & 0 & 11/41 & 0 & 11/22 & 0 & 26/61  & 0 & 21/69 \\

2 & YLDP13YL & UC & 2$^{\ast}$ & \textbf{35}/\textbf{76} & UC & 1$^{\ast}$ & 20/24 & NA & NA/NA & 0 & 8/18 & 0 & 8/42 & 0 & 8/22 &  0 & 18/62 & 0 & 22/70 \\

3 & A60 & DoS & 1 & \textbf{28}/\textbf{41} & / & 0 & 18/22 & 0 & 5/16 & 0 & 7/13 & 0 & 8/33 & 0 & 5/21 & 0 & 22/36 & 0 & 20/39 \\

4 & Mini C & / & 0 & \textbf{46}/\textbf{72} & / & 0 & 18/31 & 0 & 7/15 & 0 & 5/11 & 0 & 6/31 & 0 & 5/21 & 0 & 18/68 & 0 & 18/70 \\

5 & BR30 & / & 0 & \textbf{28/51} & / & 0 & 8/19 & NA & NA/NA & 0 & 4/11 & 0 & 4/31 & 0 & 4/20 & 0 & 13/40 & 0 & 13/48  \\

6 & Hue & / & 0 & \textbf{65}/\textbf{110} & / & 0 & 29/36 & 0 & 4/11 & 0 & 7/11 & 0 & 9/31 & 0 & 7/25 & 0 & 34/110 & 0 & 22/99 \\

7 & Base Station & / & 0 & \textbf{34}/\textbf{51} & / & 0 & 29/33 & 0 & 7/16 & 0 & 6/9 & 0 & 9/17 & 0 & 7/13 & 0 & 19/38 & 0 & 23/50 \\

8 & HS100 & NPD & 3 & \textbf{24}/64 & / & 0 & 20/27 & NA & NA/NA & 0 & 6/13 & 0 & 6/31 & 0 & 6/22 & 0 & 20/64 & 0 & 19/\textbf{71} \\

9 & HS110 & NPD & 4 & \textbf{24}/\textbf{79} & / & 0 & 17/22 & NA & NA/NA & 0 & 6/14 & 0 & 9/33 & 0 & 6/22 & 0 & 20/62 & 0 & 19/78  \\

10 & F7C027au & / & 0 & \textbf{13}/\textbf{21} & / & 0 & 7/10 & 0 & 6/14 & 0 & 8/12 & 0 & 6/18 & 0 & 6/15 & 0 & 8/14 & 0 & 12/21 \\

11 & MSS310 & / & 0 & \textbf{42}/\textbf{61} & / & 0 & 15/17 & 0 & 8/16 & 0 & 5/11 & 0 & 8/45 & 0 & 8/21 & 0 & 30/59 & 0 & 20/61 \\

12 & B25AUS & / & 0 & \textbf{19}/\textbf{42} & / & 0 & 8/13 & 0 & 7/19 & 0 & 7/14 & 0 & 11/17 & 0 & 7/11 & 0 & 16/36 & 0 & 9/41 \\

13 & Mini US & / & 0 & \textbf{25/61} & / & 0 & 8/41 & NA & NA/NA & 0 & 7/16 & 0 & 7/35 & 0 & 7/22 & 0 & 9/55 & 0 & 8/49 \\

14 & SP4L-AU & / & 0 & \textbf{37/43} & / & 0 & 18/32 & 0 & 5/11 & 0 & 5/17 & 0 & 7/32 & 0 & 5/23 & 0 & 23/40 & 0 & 17/40 \\

15 & R6400 & / & 0 & 11/37 & / & 0 & \textbf{20}/24 & 0 & 4/13 & 0 & 3/12 & 0 & 4/24 & 0 & 4/18 & 0 & 6/30 & 0 & 6/\textbf{41} \\

16 & WL100 & / & 0 & \textbf{53}/\textbf{81} & / & 0 & 38/44 & NA & NA/NA  & 0 & 8/16 & 0 & 8/46 & 0 & 8/27 & 0 & 41/70 & 0 & 29/76 \\

17 & Alro Pro 2 & / & 0 & \textbf{25}/36 & / & 0 & 16/22 & 0 & 10/14 & 0 & 8/13 & 0 & 14/22 & 0 & 10/17 &0 & 18/22 & 0 & 13/\textbf{41} \\

18 & F19821W & / & 0 & \textbf{39}/75 & / & 0 & 36/33 & 0 & 7/13 & 0 & 5/11 & 0 & 7/23 & 0 & 7/14 & 0 & 27/65 & 0 & 21/\textbf{76} \\

19 & T-131P & / & 0 & \textbf{36}/80 & / & 0 & 9/22 & 0 & 7/16 & 0 & 7/20 & 0 & 9/42 & 0 & 7/35 & 0 & 21/65 & 0 & 20/\textbf{91} \\

20 & RM mini 3 & / & 0 & \textbf{14/36} & / & 0 & 9/30 & NA & NA/NA & 0 & 10/17 & 0 & 14/31 & 0 & 10/23 & 0 & 6/30 & 0 & 5/35 \\
\bottomrule
\end{tabular}
}
\resizebox{\linewidth}{!}{
\begin{tabular}{l}
\footnotesize
\begin{tabular}[l]{@{}l@{}}\textbf{UC}: Unknown crash. \textbf{NPD}: Null pointer dereference. \textbf{DoS}: Denial of service. \textbf{T}: Vulnerability type. \textbf{C}: Number of crashes. \textbf{10/24}: Number of response categories (10 minutes/24 hours).\\$^{\ast}$: Remotely exploitable. \textbf{NA}: Since \textsc{Doona} is only applicable to some network protocols, devices that cannot be tested are represented by `NA'.\end{tabular}
\end{tabular}
}
\end{table*}

\vspace{1mm}
\noindent \textbf{Null pointer dereferences.} As shown in Table~\ref{tab:result}, the 7 crashes triggered by \textsc{Snipuzz} in TP-Link HS110 and HS100 are all caused by null pointer dereferences. 
After sending the test cases to HS110 and HS100, the devices crashed, unable to reply to any interaction. However, after a few minutes, the devices automatically restarted and recovered to the initial state.
Based on the analysis of test cases, we found that the vulnerabilities are all triggered by messages that mutated in JSON syntax. 
Put differently, when some important placeholders, such as curly braces and colons, or a part of the test message are mutated, the syntax structure and the semantic meaning of the message are broken. If the device cannot handle the mutated input message properly, it will crash the device. 
We reported the vulnerabilities to the device vendor, TP-Link, via email on June 13, 2020. They have confirmed the vulnerability and promised to fix it through a firmware update.

\vspace{1mm}
\noindent\textbf{Denial of service.} Another interesting finding is the denial of service vulnerability detected in Philips A60 smart bulb. After being tested by \textsc{Snipuzz} for 24 hours, Philips' official companion app could not manage the device normally. Specifically, the device cannot be found in the app and if any further messages are sent through the app, the response in the app will keep asking to bound the device to a device group and no further interaction is available. However, we observe that if the message packet is sent directly to the device, the device can work normally. This indicates that the device does not completely crash but its service via the companion app is denied. 

\vspace{1mm}
\noindent\textbf{Unknown crashes.} \textsc{Snipuzz} found 5 crashes on Yeelight bulbs, YLDP05YL, and YLDP13YL. The devices crashed and restarted by themselves within roughly one minute. By analyzing the test cases, we found that the crashes are due to the deletion of certain data domains, such as the nullify of parameters, marked as red in Table~\ref{tab:mutated}. As the firmware of the 2 devices is not publicly available, the root cause of the vulnerability cannot be determined; However, we can still deduce that the vulnerability is due to the device reading in null values during the parsing process, causing a crash during the assignment.
We also find that communication using a local network does not require any authentication, which means that the device can be crashed by any attackers in the local network. Therefore, we consider the vulnerabilities as `remotely exploitable'.

\begin{table}[t]
\centering
\caption{Mutated messages of \textsc{Snipuzz} \& \textsc{IoTFuzzer}.}
\vspace{-3mm}
\small
\resizebox{0.9\linewidth}{!}{
\begin{tabular}{@{}ll@{}}
\toprule
\textbf{Contents of mutated messages} & \textbf{Generated by} \\ \midrule

\begin{tabular}[c]{@{}l@{}}\{"\{"id": 0, "method": "start\_cf", "params": ["4, 4,  "1000,\\~ 2, 2700,100,500
 ,1,255,10,5000,7,0,0,500,2,5000,1"]\}"\end{tabular} &  Original Message \\ \midrule

\begin{tabular}[c]{@{}l@{}}\{"\{"id": 0, "method": "start\_cf", "params": ["\textcolor{red}{4, ,}  "1000,\\~ 2, 2700,100,500
 ,1,255,10,5000,7,0,0,500,2,5000,1"]\}"\end{tabular} &  \textsc{Snipuzz} \\ \midrule
 
\begin{tabular}[c]{@{}l@{}}\{"\{"id": 0, "method": "start\_cf", "params": ["\textcolor{red}{ , 4,} "1000,\\~  2, \textcolor{red}{270000},100,500
 ,1,255,10,5000,7,0,0,500,2,5000,1"]\}"\end{tabular}&  \textsc{IoTFuzzer} \\ \midrule
\end{tabular}
}
\label{tab:mutated}
\end{table}

\subsubsection{Benchmark with state-of-the-art tools}
As shown in Table~\ref{tab:result}, for 24 hours fuzz testing on each devices, none of the benchmark tools found a crash except for \textsc{IoTFuzzer}. They did not find the crash due to various reasons.
\textsc{Donna} focuses more on the mutation of communication protocols. Further, \textsc{Donna} cannot be applied on all devices, which also limits its capacity. 
Since \textsc{Boofuzz} directly replaces the specified positions in the message with a preset string, it can only trigger a limited types of vulnerabilities. 
\textsc{Nemesys} offers a new idea of determining message snippets. However, since it determines message snippets by the distribution of values in messages, it is difficult for \textsc{Nemesys} to accurately decide the boundary between data and non-data domains.
Therefore, \textsc{Nemesys} can hardly detect vulnerabilities that can only be triggered by mutating the data or non-data domains.
\textsc{Snipuzz-NoSnippet}, which does not apply the snippet-based mutation method used in \textsc{Snipuzz}, is similar to the classic fuzzer AFL\cite{AFL}. Since \textsc{Snipuzz-NoSnippet} does not infer the structure of the message but directly uses single or multiple consecutive bytes as the unit of mutation, most of the test cases generated by \textsc{Snipuzz-NoSnippet} destroy the structure of the messages. Such a method is difficult to work on devices that require highly-structured inputs.

\textsc{IoTFuzzer} detected 2 crashes in 2 smart bulb devices, \ie, the YLDP05Y and YLDP013Y. 
Due to the mutation strategy of \textsc{IoTFuzzer}, the malformed input provided by \textsc{IoTFuzzer} is obtained by emptying the data domain.
According to the mutated messages listed in Table~\ref{tab:mutated}, we can see that the messages mutated by \textsc{IoTFuzzer} resemble the ones generated by \textsc{Snipuzz}.
The mutated domains of messages from \textsc{Snipuzz} and \textsc{IoTFuzzer} in Table~\ref{tab:mutated} are all in the data domain. 
In terms of the effect of the mutation test, \textsc{Snipuzz} and \textsc{IoTFuzzer} achieve the same goal on these two messages.
However, \textsc{Snipuzz} can cover the mutation space of \textsc{IoTFuzzer} because \textsc{IoTFuzzer} only focuses on the data domain mutation while \textsc{Snipuzz} can mutate both the data and non-data domains.

To further determine the root cause of the crash, we obtained the firmware source code of HS100 and HS110, two typical market consumer-grade smart plugs manufactured by TP-Link, and conducted a case study which reflected the differences between \textsc{Snipuzz} and \textsc{IoTFuzzer}. We found that one of the crashes triggered by \textsc{Snipuzz} on the two devices is caused by breaking the syntax structure and  mutating both on data and non-data domains. More specifically, the mutated messages successfully bypassed the sanitizer and triggered the crash during function execution. We deduce that this could be caused by an error-prone third-party sanitizer (more details could be found in Appendix~\ref{sec:case_study}). On the other hand, due to the design of \textsc{IoTFuzzer}, the fuzzing is based on the grammatical rules as the \textsc{IoTFuzzer} tends to satisfy the grammar requirements with first-priority, in order not to be rejected by the sanitizer and ensure that each test case can reach the functional execution part in the firmware. Such strategy constraints the test range of fuzzing and its capacity to cover the sanitization part in comparison to \textsc{Snipuzz}. Therefore, we argue that considering the complexity of IoT firmware testing, a lightweight and effective black-box vulnerability detection tool, such as \textsc{Snipuzz}, is a pressing need.

\subsection{Runtime Performance} \label{sec:runtime_perf}

\begin{figure}[t]
\centering
\includegraphics[width=\linewidth]{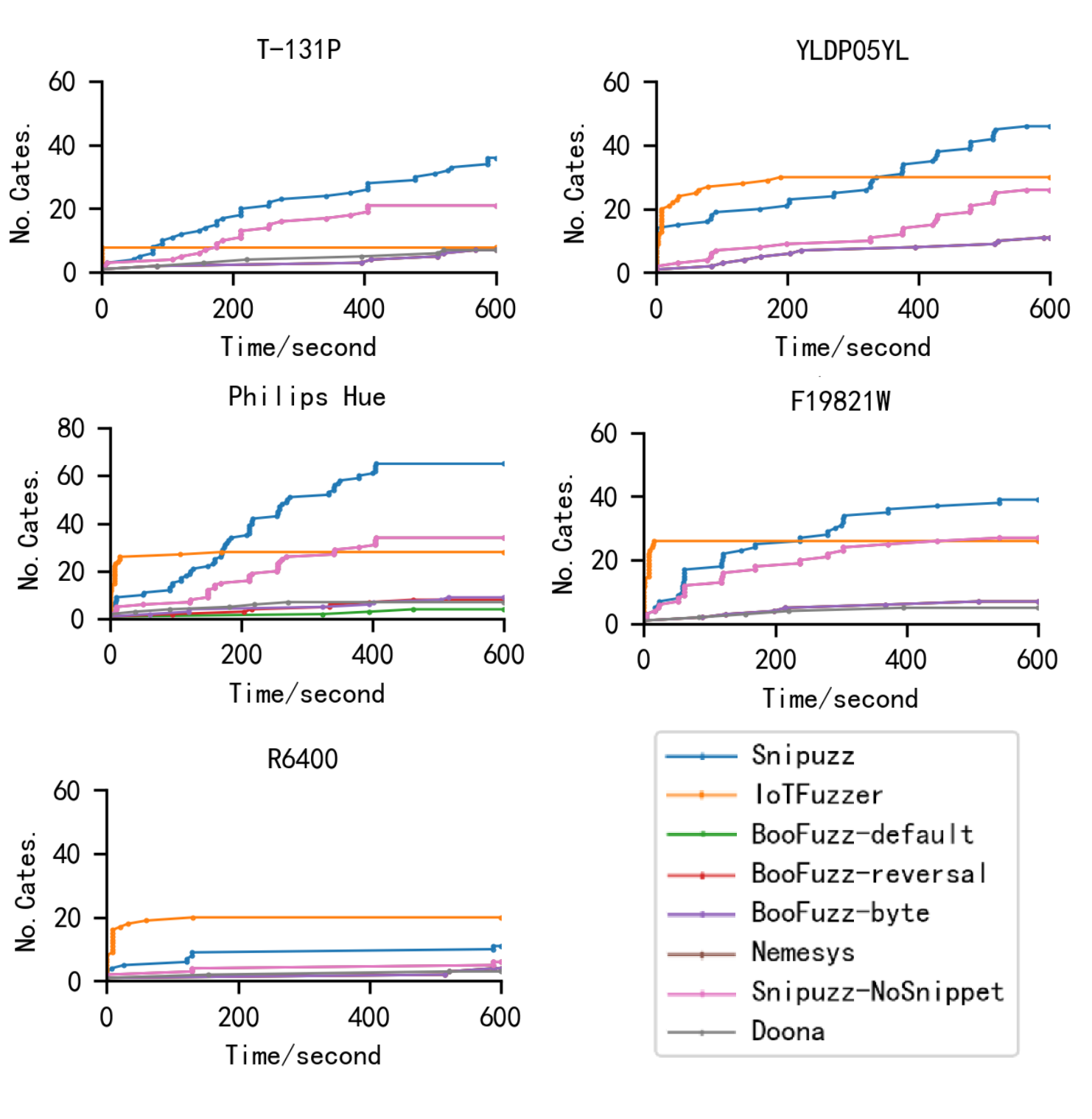}
\caption{\small The number of categories discovered over time.}\label{evaluation1}
\end{figure} 

Figure~\ref{evaluation1} shows how \textsc{Snipuzz} and the other seven fuzzers explored the device firmware during the first \textbf{10 minutes}. Limited by spacing, we only present the results of 5 devices here but plot results of all 20 devices in Appendix~\ref{appendix_runtime}.
We repeated the fuzz testing for 10 times and recorded the medium values of the numbers of response categories discovered by each method, indicating that the coverage has been explored. We manually review the presented response categories to remove the mis-categorization caused by randomness in responses or the response mechanism of devices.

As shown in Figure~\ref{evaluation1}, \textsc{Doona} can only detect a small number of response categories. \textsc{Doona} is protocol-based fuzzing methods, and its tests are more biased towards protocol content. The mutation test on the communication protocol has a high probability of being directly rejected or ignored by the device unilaterally, resulting in few categories of responses that can be received. 

We implemented three fuzzing strategies based on \textsc{Boofuzz}, \ie, mutating the whole message as a string, mutating each byte of the message, and mutating non-data domain. However, the testing results indicate that all of them explored very limited categories of responses on each device. The limitation of category discovery is due to the mutation strategy of \textsc{Boofuzz}, which replaces the target contents with a specific pre-defined string. For example, using strings, such as \texttt{``/./././././././.''}, to replace the content of messages in different strategies (\eg, replacement of the entire strings, a single byte, or a non-data domain), causes the violation of message format and could be easily rejected by the sanitizer. Therefore, most of the responses obtained by \textsc{Boofuzz} fall into the category of ``error responses''.

The number of response categories explored by \textsc{IotFuzzer} grows rapidly within a short period of time and then stagnates. 
In the mutation stage, \textsc{IotFuzzer} randomly selects a set of inputs from the original candidate inputs and randomly mutates the data domain for one or more message(s). It will continue to repeat this method until the device crashes or reaches the time limit. Such a method based on randomness helps \textsc{IotFuzzer} to mutate and test a large number of message data domains in the original input and collect response message categories quickly in the beginning. However, the number of response categories found by \textsc{IotFuzzer} will soon reaches the limitation due to the data domain mutation. 

In most devices, \textsc{Snipuzz} has maintained a steady upward trend in most cases, and after a period of time the number of response categories found by \textsc{Snipuzz} exceeds \textsc{IotFuzzer}. Unlike \textsc{IotFuzzer}, \textsc{Snipuzz} mainly searches for the response categories through the \textbf{Snippet Determination} stage. As per the message snippet exploration strategy, \textsc{Snipuzz} first explores all the response categories of a certain message as many as possible. After the snippets of a message are obtained and tested by \textbf{Snippet Mutation}, the next message will be processed in the same way until all messages in the initial message sequence have been tested. Followed by this method, \textsc{Snipuzz} may not get a large number of response categories in a short time.
When \textsc{Snipuzz} detects a message snippet, every byte in the message content will be included in the test. Therefore, as shown by the bold numbers in Table~\ref{tab:result}, for 15 out of 20 devices, \textsc{Snipuzz} covers the most number of response categories after 24-hour fuzz testing, compared to other state-of-the-art IoT fuzzing tools. 

On 5 devices, \textsc{Snipuzz-NoSnippet} collected more response categories than \textsc{Snipuzz} within 24 hours. The mutation method used by \textsc{Snipuzz-NoSnippet} is similar to the classic fuzzer AFL~\cite{AFL}. It directly performs mutation on a single byte or several consecutive bytes. 
However, \textsc{Snipuzz-NoSnippet} is difficult to cover response categories that are not obtained by breaking the grammatical format (\eg, data out of bounds in the data domain). 
Theoretically, although the \textsc{Snipuzz-NoSnippet} mutation method is not so efficient, it still has the capability to explore the most categories of responses.

\textsc{Nemesys} explores more categories of responses than \textsc{BooFuzz} and \textsc{Doona}, but does not exceed \textsc{Snipuzz}. The \textsc{Nemesys} strategy performs deterministic mutations on each data domain of the messages in turn, which makes its trend of run-time performance similar to \textsc{Snipuzz}.
However, the data domain determination strategy of \textsc{Nemesys} is not based on the responses from IoT device. Thus, the distribution of byte values in messages does not benefit in covering more response categories. Therefore, the number of response categories collected by \textsc{Nemesys} is limited.

It is interesting to observe that, in the case of R6400, \textsc{Snipuzz} also enters a stagnation after only finding a few response categories. We carefully checked the initial input message sequences and found that the average length of the message exceeds 400 bytes, forcing \textsc{Snipuzz} to generate and send a large number of probe messages to determine message snippets.
Therefore, in the first 10 minutes, \textsc{Snipuzz} was still exploring the response category of the first few messages, so it did not exceed \textsc{IotFuzzer}.

\subsection{Assessment on Message Snippet Inference}\label{sec:assessment_snippet_inference}

Among all strategies, \textsc{Snipuzz} and \textsc{Nemesys} utilize semantic segmentation, to assess their performance of message snippet inference. We compare the snippets they produce during the fuzzing process with the grammar rules defined in API documents. Specifically, for some mature and popular languages, such as JSON, we establish the grammar rules as per their standard syntax; for custom formats, such as strings or custom bytes, we refer to the official API documents and define the grammar rules based on the instructions.

Equation~(\ref{equ:distance}) quantifies the quality of snippet inference, and \textit{Similarity} indicates the percentage of correctly categorized bytes in a snippet-determined message, $m$, compared with the ground truth, $g$, manually extracted from the grammar rules. 

\begin{equation}
\begin{split}
    Similarity(m) = 1 - \frac{count[cate(m) \oplus cate(g)]}{len(m)},
\end{split}
\label{equ:distance}
\end{equation}

\noindent where $cate()$ returns the category of each message byte in a series of ``0'' and ``1'' bits, $count()$ counts the number of mis-categorized bytes, and $len()$ represents the length of a message. Note that in a ground truth message, ``0'' indicates the non-data domain (marked blue in Table~\ref{tab:case2}), while ``1'' indicates the data domain (marked red in Table~\ref{tab:case2}). Therefore, the $\oplus$ is the bitwise $XOR$ operation.

\begin{table}[t]
\centering
\caption{Inference results of \textsc{Snipuzz} and \textsc{Nemesys}.}\label{tab:case2}
\vspace{-3mm}
\small
\resizebox{0.95\linewidth}{!}{
\begin{tabular}{@{}ll@{\quad}l}
\toprule
\textbf{Method} & \textbf{Ave. Similarity}& \textbf{Example} \\ \midrule

 \textsc{Snipuzz} & 87.1\%
 &\textcolor{blue}{\{"} \textcolor{red}{on} \textcolor{blue}{":true,"} \textcolor{red}{sta} \textcolor{blue}{":140,"}  \textcolor{red}{bri} \textcolor{blue}{":254\}} \\ 

\textsc{Nemesys} & 64.5\%
 &\textcolor{blue}{\{"on":} \textcolor{red}{true} \textcolor{blue}{,"sta":} \textcolor{red}{140} \textcolor{blue}{,"bri"}  \textcolor{red}{254\}} \\ 

Grammar & 100.0\%
&\textcolor{blue}{\{"} \textcolor{red}{on} \textcolor{blue}{":} \textcolor{red}{true} \textcolor{blue}{,"}  \textcolor{red}{sta} \textcolor{blue}{":}  \textcolor{red}{140}
\textcolor{blue}{,"} \textcolor{red}{bri} \textcolor{blue}{":} \textcolor{red}{254}
\textcolor{blue}{\}}
\\ \midrule

\end{tabular}
}
\end{table}

In addition, followed by Equation~(\ref{equ:distance}), we compute the average similarity of the snippets (or data domain) determined by \textsc{Snipuzz} and \textsc{Nemesys} for all the 235 messages obtained from experiments. 
Note that during the calculation of the average similarity, for each message, if there are multiple snippet sets determined, we will select the snippet inference with the highest similarity value; therefore  a snippet could reflect the grammatical rules as many as possible and maximize the performance of message semantic segmentation.

The average similarity result of \textsc{Snipuzz}, 87.1\%, indicates that, by applying snippet inference based on the hierarchical clustering approach, \textsc{Snipuzz} can effectively find the grammatical rules hidden in the message. Ideally, in \textsc{Snipuzz}, the merging of clusters removes the influence caused by the randomness in responses and by the replying message mechanism itself. Therefore, the message snippets will conform to the grammatical rules gradually, which leads \textsc{Snipuzz} to a higher similarity result.

However, we also found some differences between the snippet inference method and the grammatical rules in some results. For example, given the example shown in Table~\ref{tab:case2}, the snippet inference method combines the strings belonging to the data domain in the grammatical rules (\ie, `true', `140' and `254') with some placeholders (such as double quotes and curly brackets). After analyzing the response messages, we found that the responses obtained after destroying these data domains and destroying placeholders are all about invalid format. 
This may be due to the fact that in the firmware, when an error occurs in the parsing format, the response does not report a detailed description of the error but instead returns a general format error. 

On the other hand, \textsc{Nemesys} uses the distribution of value changes in the protocol to determine the boundary of different data domains, and to achieve the semantic segmentation of a message. 
The advantage of this method is that it does not require any other additional information, such as grammar rules or a large number of training data sets in addition to the message itself.

The average similarity result of \textsc{Nemesys}, 64.5\%, is lower than the \textsc{Snipuzz} result. 
Given the example shown in Table~\ref{tab:case2}, when segmenting messages in a format requires restricted syntax, such as Json and XML, \textsc{Nemesys} can achieve a good semantic segmentation performance, because the placeholders usually use symbols unusually used in data domains. This distribution of byte value enables \textsc{Nemesys} to effectively find the boundaries between data domains. However, in IoT devices, customized formats are prevalent. For example, the smart bulb BR30 uses custom bytes as a means of communication, where each byte corresponds to a special meaning (\ie, \textit{"0x61"} represents \textit{"CHANGE\_MODE"} and \textit{"0x0f"} represents \textit{"TRUE"}). In such cases, the value distribution of characters can no longer be used as a guidance for the data domain determination, and thus the message segmentation determined by \textsc{Nemesys} is error-prone.

\section{Discussion and Limitations} \label{appendix_limit}

\textsc{Snipuzz} has successfully examined 20 different devices and exposed security vulnerabilities on five of them. However, there are still some limitations relevant to efficiency and scalability of \textsc{Snipuzz}. 
We discuss the limitations in this section and propose solutions as future work.

\vspace{1mm}
\noindent \textbf{Scalability and manual effort.} IoT devices can be tested by \textsc{Snipuzz} if the valid network packets are known. In our prototype, we capture communication packets by running API programs and monitoring network communication (Note that packets can also be obtained by statically analyzing API programs without running them).
In the absence of API programs or documents, we can recover the message formats from the official Apps of IoT devices through decompilation and taint analysis. Or as a second way, we can solve this problem by intercepting the communication between APPs and IoT devices, and then recovering message formats from the captured packets. The second way is feasible and we have experimented it in TP-Link’s IoT control APP KASA, which can be further developed for more IoT devices. However, both methods could introduce overhead and involve manual effort.

Recall in Section~\ref{sec:initial_seed_acquisition} that \textsc{Snipuzz} requires manual effort, which takes 5 man-hours per device to collect the initial seeds during the message sequence acquisition phase. The manual effort is mainly referred to cleaning the packets from the API programs that are obtained from publicly available first- and third-party resources. 
To mitigate this limitation when applying \textsc{Snipuzz} to IoT devices, techniques such as crawlers could be used to automatically gather API programs associated with the IoT devices in the future work. Moreover, the process of cleaning the packets could also be improved by pre-processing keywords through scripts to achieve automatic collection of communication packages.

\vspace{1mm}
\noindent \textbf{Threats to validity.} As \textsc{Snipuzz} collects initial message sequences via API programs and network sniffers, the first threat comes from the absence of API programs. In this case, we can recover message formats based on the companion apps of IoT devices (similar to \textsc{IoTFuzzer}) but may need more manual efforts. Second, the encryption in messages decreases the effectiveness of snippet determination because the semantic information could be corrupted. A potential solution to the encryption issue is to integrate decryption modules into \textsc{Snipuzz}. Finally, the code coverage of firmware could be subject to the accessibility of API programs, since \textsc{Snipuzz} can only examine the functionalities that are covered in API programs. Recombining the message snippets from different seeds to generate new valid inputs could mitigate this limitation.

\vspace{1mm}
\noindent \textbf{Encryption.} 
During Message Acquisition, we noticed that encryption is used to protect communication in some API programs. 
Encryption has no effect on the message sequence mutation process, but the snippet determination process basically fails. Because the encryption algorithm disrupts the original format of the message, the segmentation of snippets is sensitive to the position of the character. Moreover, because the response messages from the device are also encrypted, \textsc{Snipuzz} cannot get useful feedback from them. Similarly, the encryption and decryption algorithms in the API program can be integrated into the \textsc{Snipuzz} module to address this limitation, or the difficulties caused by encryption can be addressed from the perspective of mutation strategy design.

\vspace{1mm}
\noindent \textbf{Coverage.} 
The code coverage of firmware explored by \textsc{Snipuzz} depends on the API programs. For example, if the API programs of a bulb only support the functionality of turning on power, it is almost impossible to explore the functionality of adjusting the brightness via mutating the messages captured during the power turned on. 
In the future work, without the support of grammar, we will consider recombining the message snippets to try to generate new valid inputs. This method can help explore more firmware execution coverage in addition to the original inputs provided.

\vspace{1mm}
\noindent \textbf{Requirements on detailed responses.}
The detection effectiveness of \textsc{Snipuzz} depends on the quality of message snippets which is contingent on how much information could be obtained from the responses of IoT devices. To put differently, if the IoT device does not provide responses that are detailed enough, for example 
reporting all the errors with a uniform message, it could be hard for \textsc{Snipuzz} to determine the message snippets. Fortunately, in many IoT devices, advanced error descriptions could be obtained in debug mode which will significantly improve the determination process of message snippets in \textsc{Snipuzz}.

\vspace{4mm}

\section{Related Work}

Our \textsc{Snipuzz} performs in a black-box manner for detecting vulnerabilities in IoT devices. Unlike existing black-box fuzzing for IoT devices, which blindly mutates messages, \textsc{Snipuzz} optimizes the mutation process of black-box fuzzing via utilizing responses. This feedback mechanism improves the effectiveness of bug discovery. For instance, \textsc{IoTFuzzer} \cite{Chen2018IOTFUZZER} obtains the data 
domain, on which \textsc{IoTFuzzer} performs blind mutation. Thus, \textsc{IoTFuzzer} lacks the knowledge of the quality of the generated inputs, resulting in a waste of resource on the low-quality inputs. There are also several dynamic analysis approaches focusing on the networking modules of IoT devices. For example, SPFuzz defines a new language for describing protocol specifications, protocol state transitions, and their correlations~\cite{song2019spfuzz}. SPFuzz can ensure the correctness of the message format in the conversation state and the dependence of the protocol. IoTHunter is a grey-box approach to fuzz the state protocol of IoT firmware~\cite{yu2019poster}. IoTHunter can constantly switch the protocol state to perform a feedback-based exploration of IoT devices. In a recent example, \textsc{AFLnet} acts as a client and continuously replays the variation of the original message sequence sent to target (\ie, server or device)~\cite{Pham2020AFLNET}. \textsc{AFLnet} uses response codes, which are the numbers indicating the execution states, to identify the execution status of targets and explore more regions of their networking modules.

Another research line for dynamic analysis of IoT devices is the usage of emulators. The disadvantages of emulation are the heavy engineering efforts and the requisite of firmware, although the emulation of IoT firmware can analyze more thoroughly than black-box fuzzing. Two major challenges for emulation of IoT firmware are the scalability and throughput. Therefore, the efforts in improving the performance of emulation include full-system emulation~\cite{Muench2018What, Chen2016Towards}, improvement of emulation success rates~\cite{kim2020firmae}, hardware-independent emulation~\cite{feng2020p2im, srivastava2019firmfuzz}, and combination of user- and system-mode emulation~\cite{Zheng2019FIRM}. Based on the emulation, fuzzing can be integrated into those frameworks and can hunter defects in firmware~\cite{srivastava2019firmfuzz, Zheng2019FIRM}. 

Static analysis of firmware is the complementary approach of dynamic analysis. Semantic similarity is one of the major techniques that make static analysis successful. Researchers analyze semantic similarity via comparison of files and modules~\cite{Costin2014Large}, Control Flow Graphs (CFGs)~\cite{dullien2005graph}, parser and complex processing logic~\cite{Cojocar2015PIE}, and multi-binary interactions~\cite{redini2020karonte}. There are also many similarity-based approaches that can detect vulnerabilities across different firmware architectures. They usually extract various architecture-independent features from firmware for each node in a CFG to represent a function, and then check whether two functions' CFG representations are similar~\cite{Eschweiler2016discovRE, Pewny2015Cross}.

\vspace{2mm}
\section{Conclusion}

In this paper we have presented a black-box fuzzing framework \textsc{Snipuzz} designed for detecting vulnerabilities hiding in IoT devices. 
Different from other black-box network fuzz testing, \textsc{Snipuzz} uses the response messages returned by the device to establish a feedback mechanism for guiding the fuzzing mutation process. In addition, \textsc{Snipuzz} infers the grammatical role of each byte in the messages based on the responses from the device, so that \textsc{Snipuzz} can generate test cases that meet the device's grammar without the guidance of grammatical rules. 
We have used 20 consumer-grade IoT devices from the market to test \textsc{Snipuzz}, and it has successfully found 5 zero-day vulnerabilities on 5 different devices.

\newpage
\bibliographystyle{ACM-Reference-Format}
\bibliography{references}

\appendix
\section*{Appendix}

\begin{figure*}[t]
\centering
\includegraphics[width=\linewidth]{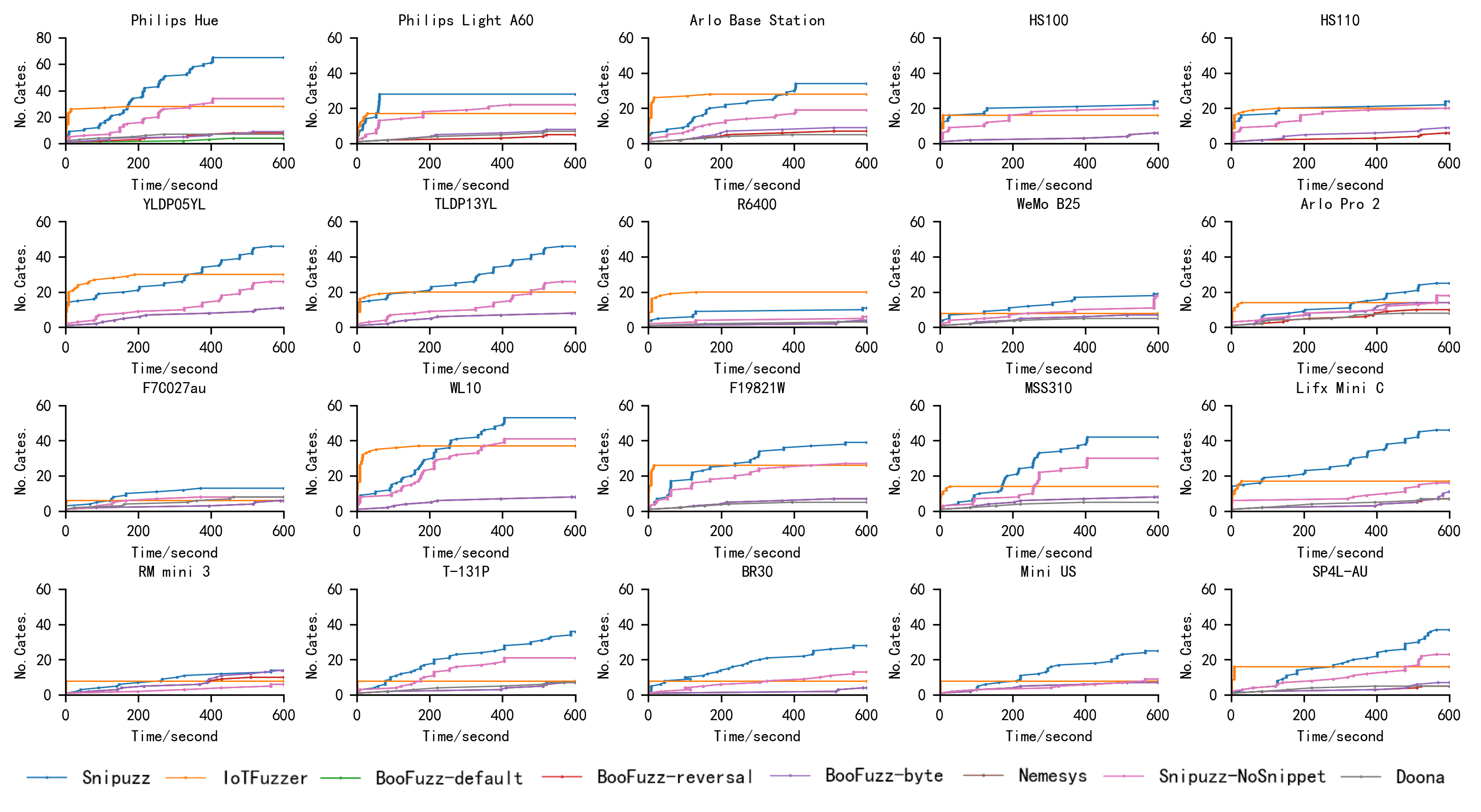}
\caption{Runtime performance. The number of categories discovered in 10 minutes on all the 20 IoT devices. \textsc{Snipuzz} performs the best on 19 devices.}\label{fig:run-time_performance_full}
\end{figure*}

\section{Runtime Performance}\label{appendix_runtime}

Fig~\ref{fig:run-time_performance_full} shows the run-time performance of \textsc{Snipuzz} and other seven baselines during the first \textbf{10 minutes}. In most benchmarks, \textsc{Snipuzz} discovers the most number of categories. Since \textsc{Snipuzz} spends time on snippet determination, it discovers less categories than \textsc{IoTFuzzer} in the beginning. However, \textsc{IoTFuzzer} quickly reaches its peak  and cannot discover new categories. On the contrary, after the stage of snippet determination, \textsc{Snipuzz} gradually discovers more categories than \textsc{IoTFuzzer} and other baselines. More detailed analysis can be found in Section~\ref{sec:runtime_perf}.

\section{Mutation Effectiveness: A Case Study}\label{sec:case_study}

The HS100 and HS110 manufactured by TP-Link are 2 classic market consumer-grade smart plugs. In the work by Chen \etal~\cite{Chen2018IOTFUZZER}, they use HS110 with firmware version 1.3.1 to test \textsc{IoTFuzzer}. The results of their experiment show that \textsc{IoTFuzzer} triggered a vulnerability in the device by mutating the data domain in a message (changing \underline{\textbf{``light''}} to \underline{\textbf{0}}). 

\begin{figure}[t]
    \centering
    \includegraphics[width=\linewidth]{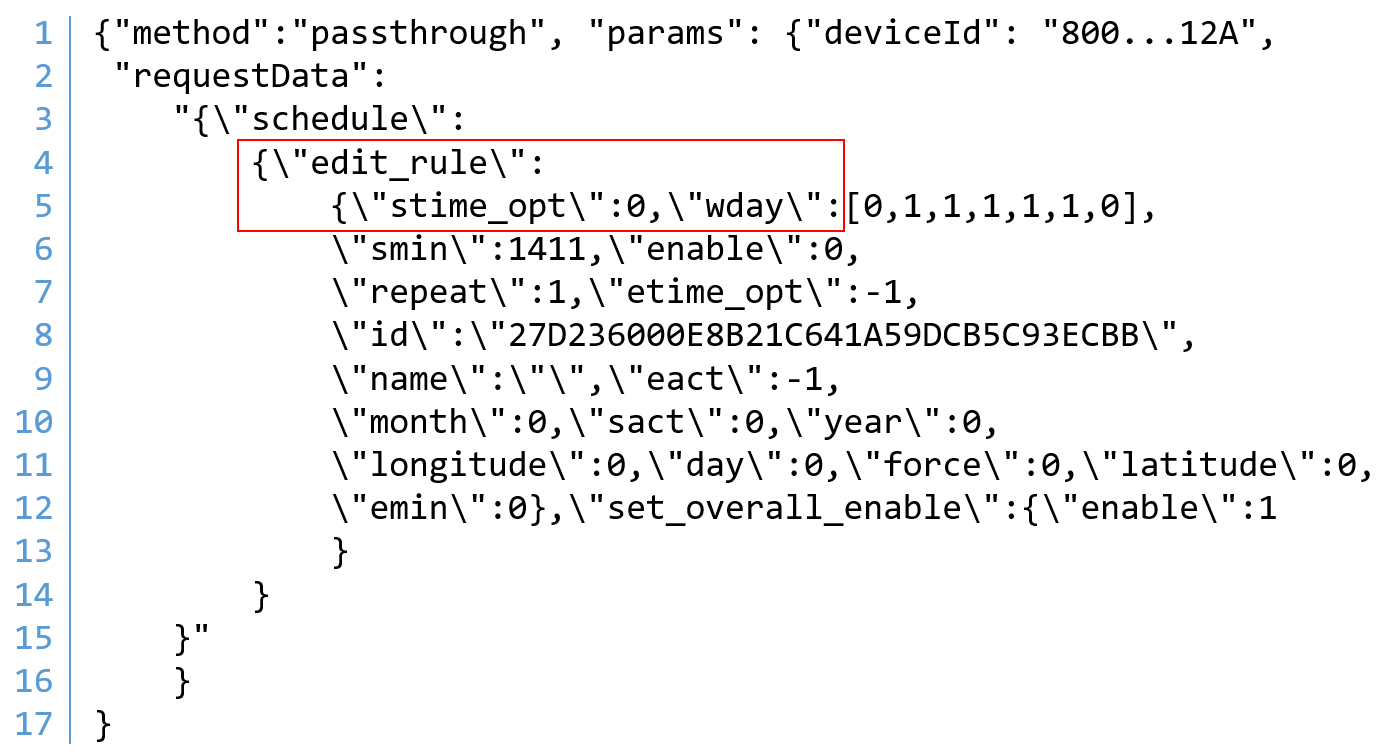}
    \vspace{-3mm}
    \caption{\small An example of vulnerability triggering}
    \label{fig:code_snippet}
    \vspace{-3mm}
\end{figure}

However, in the updated version of the firmware (1.5.2), \textsc{IoTFuzzer} did not find any vulnerabilities but \textsc{Snipuzz} did. 
Figure~\ref{fig:code_snippet} 
shows an example of the original input message and the mutated snippets (inside the red frame) in the mutated message that can trigger the vulnerability. 
In this case, \textsc{Snipuzz} triggered a vulnerability related to firmware input by breaking the JSON syntax structure in the message. The intention of the original message is to change some attributes (\eg, \texttt{`stime\_opt'} \& \texttt{`wday'}) in a rule (inferred by \texttt{`edit\_rule'}). In the mutated message, \textsc{Snipuzz} randomly deleted some contents (inside the blue frame), which break the JSON syntax. This may cause errors about parsing messages or passing parameters incorrectly handled by the firmware and, consequently, crashes the device. 

\begin{figure}[t]
    \centering
    \includegraphics[width=0.53\linewidth]{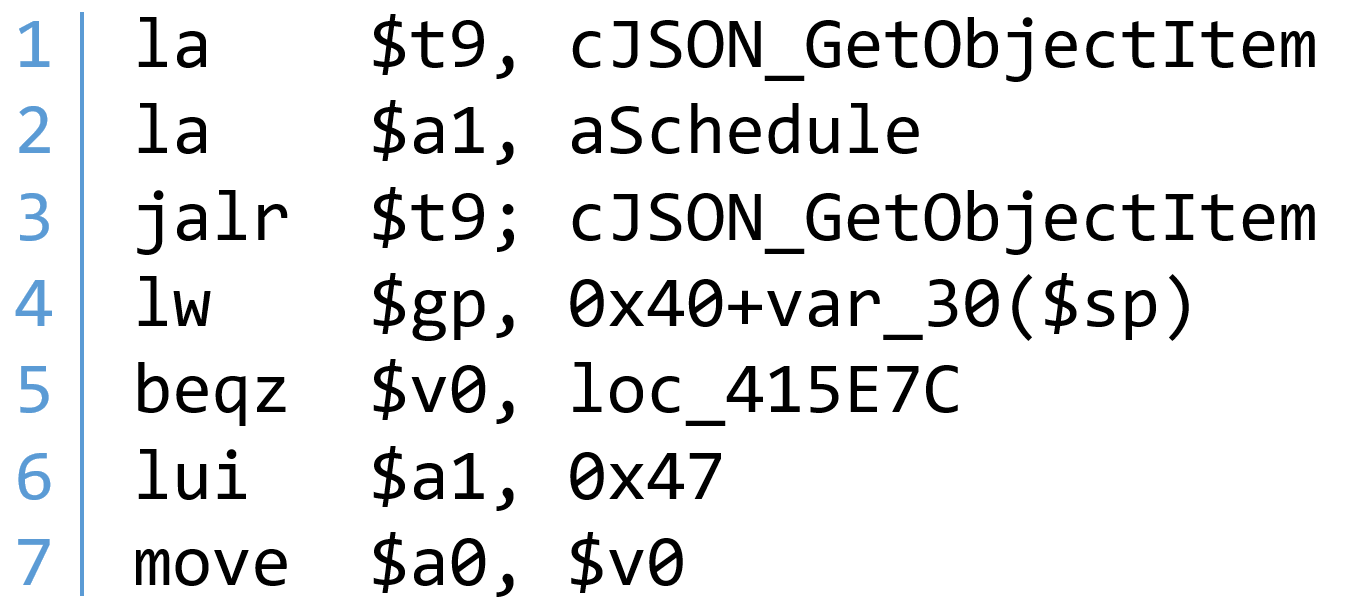}
    \vspace{-3mm}
    \caption{\small A vulnerable code snippet from HS110 firmware.}
    \label{lst:firmware}
\end{figure}

To further determine the root cause of the crash, we obtained the firmware source code. 
Figure~\ref{lst:firmware}
shows a code snippet from the firmware, using  \textsc{cJSON},\footnote{\url{https://github.com/DaveGamble/cJSON}} a popular open-source lightweight JSON parser (5.4k stars in GitHub), to interpret input message fragments. The \texttt{jalr} instruction will save the result of \texttt{cJson\_GetObjectItem} in \texttt{\$t9} and jump to this address unconditionally (see line 3 in Figure~\ref{lst:firmware}), which means the firmware will pick the value corresponding to \textit{`schedule'}. In the original message, the value corresponding to `schedule' is a JSON object headed by \texttt{`edit\_rule'} (from line 4 to line 16). 
Note that the aforementioned snippet-based mutation strategy implemented in \textsc{Snipuzz} is able to break the syntax structure and mutate both on data and non-data domains.
Interestingly, although the removing of two left curly braces breaks the JSON syntax, it is not recognized by \textsc{cJSON} parser, so the mutated message successfully bypasses the syntax validation and enters the functional code in firmware. 
When the firmware tries to access the successor JSON object in \texttt{`schedule'}, \ie, the object starts with \texttt{`edit\_rule'}, since the corresponding value is no more a JSON object, but an array, a null pointer exception is triggered.

Due to the design of \textsc{IoTFuzzer}, the fuzzing based on grammatical rules will offer priority to satisfying the grammar requirements in the mutation process in order not to be rejected by the firmware grammar detector. The advantage of this is to ensure that each test case can reach the functional execution part of the firmware. However, in this case, the test range of fuzzing based on grammatical rules cannot cover the firmware sanitising part. 

To conclude, the root cause of the crash has two factors: 1) the validation of message syntax heavily relies on a third-party library; 2) the firmware does not correctly handle the null pointer exception caused by data type mismatch.
Although it is not reasonable to require a vendor to develop products purely from scratch, we argue that thorough testing and validation on the open-source library are essential. Considering the complexity of IoT firmware testing, a lightweight and effective black-box vulnerability detection tool, such as \textsc{Snipuzz}, is a pressing need.

\end{document}